\documentclass[epj,final]{svjour}
\usepackage{graphicx}
\usepackage{amsfonts}
\usepackage{amssymb}
\usepackage{psfrag}
\usepackage{xcolor}
\usepackage{soul}

\def\CC{}
\def\AG{}

\def\CCb{}

\def\be{\begin{equation}}
\def\ee{\end{equation}}
\def\ba{\begin{eqnarray}}
\def\ea{\end{eqnarray}}

\def\trace{\mathop{\rm Tr}\nolimits}
\def\ket#1{|{#1}\rangle}
\def\bra#1{\langle{#1}|}

\begin{document}
\title{Absence of logarithmic divergence of the entanglement entropies
at the phase transitions of a 2D classical hard rod model}

\author{Christophe Chatelain${}^1$ \and Andrej Gendiar${}^2$}

\institute{${}^1$ Universit\'e de Lorraine, CNRS, LPCT, F-54000 Nancy, France\\
${}^2$ Institute of Physics, Slovak Academy of Sciences,
D\'ubravska\'a cesta 9, SK-845 11, Bratislava, Slovakia}

%\date{\today}

\abstract{
Entanglement entropy is a powerful tool to detect continuous, discontinuous
and even topological phase transitions in quantum as well as classical systems.
In this work, von Neumann and Renyi entanglement entropies are studied
numerically for classical lattice models in a square geometry. {\CC A cut
is made from the center of the square to the midpoint of one of its edges, say the
right edge. The entanglement entropies measure the entanglement between the
left and right halves of the system.} As in the strip geometry,
von Neumann and Renyi entanglement entropies diverge logarithmically at the
transition point while they display a jump for first-order phase transitions.
The analysis is extended to a classical model of non-overlapping finite hard
rods deposited on a square lattice for which Monte Carlo simulations have shown
that, when the hard rods span over 7 or more lattice sites, a nematic phase
appears in the phase diagram between two disordered phases. A new Corner Transfer Matrix
Renormalization Group algorithm (CTMRG) is introduced to study this model.
No logarithmic divergence of entanglement entropies is observed at the phase
transitions in the CTMRG  calculation discussed here.
We therefore infer that the transitions neither can belong to
the Ising universality class, as previously assumed in the literature, nor be
discontinuous.
}

\maketitle

\section{Introduction}
The quantum entanglement between the two subsystems $A$ and $B$ of
a macroscopic system has attracted a considerable interest in the
last decade~\cite{Gu,Chen,Amico}. Besides its purely theoretical
interest, the entropy that quantifies this entanglement have found
some applications, in particular in the identification of phase
boundaries as will be discussed in this work. Denoting
    \be\rho_A=\trace_B\ket{\psi_0}\bra{\psi_0}\ee
the reduced density matrix of subsystem $A$ in the ground
state $\ket{\psi_0}$ of the system, the von Neumann entanglement
entropy of the degrees of freedom of $A$ with those of subsystem
$B$ is defined as
	\be S_A=-\trace\rho_A\log\rho_A\ee
while the Renyi entropies are
	\be S_n={1\over 1-n}\log \trace\rho_A^n.\ee
In dimension $1+1$ and with Open Boundary Conditions, Conformal Field Theory
predicts that von Neumann entanglement entropy diverges logarithmically when
approaching a critical point~\cite{Calabrese}
    \be S_A={c\over 6}\ln{\xi\over a}+c'\ee
where the correlation length $\xi$ scales with the control parameter $\delta$
as $\xi\sim|\delta|^{-\nu}$. The prefactor is proportional to the central
charge $c$ which is a universal quantity. At the critical point, the
entanglement entropy diverges as $S_A\sim {c\over 6}\ln\ell$ with the
length $\ell$ of the subsystem $A$. Similarly, Renyi entropies behave as
$S_n\sim {c\over 12}\left(1+{1\over n}\right)\ln\ell$. As observed
numerically for the quantum $q$-state Potts chain with $q>4$~\cite{Lajko},
the entanglement entropy $S_A$ displays a jump at a first-order phase
transition.
\\

{\CC Entanglement entropies $S_A$ are easily obtained in DMRG calculations
of quantum spin chains because the reduced density matrix $\rho_A$ is computed
and diagonalized at each iteration. The approach has been extended to
two-dimensional classical systems by using the eigenvector $\ket{\psi_M}$
associated to the largest eigenvalue of the transfer matrix to construct the
density matrix as $\rho=\ket{\psi_M}\bra{\psi_M}$ and then the reduced
density matrix $\rho_A$ by a partial trace. When the classical transfer
matrix can be interpreted as the evolution operator in imaginary time of a 1D
quantum Hamiltonian~\cite{Fradkin,Trotter,Suzuki1,Suzuki2,Ueda}, the entropy
$S_A=-\trace_A\rho_A\ln\rho_A$ measures the quantum entanglement between the
degrees of freedom lying in $A$ with those in $B$. By abuse of langage, one
may say that $S_A$ measures the entanglement between the left and right
part of the strip on which the classical system lives. The entanglement
entropy has proved to be a useful quantity in classical systems: the phase
diagram can be determined from the entanglement entropy, even when it involves
topological phase transitions~\cite{Chatelain}. In the CTMRG algorithm,
the reduced density matrix of a cut of width $L/2$ in a square lattice of size
$L\times L$ is constructed as $\rho_A=C^4/\trace C^4$~{\AG \cite{Krcmar1,Krcmar2}}.
Moreover, the CTMRG algorithm requires the corner transfer matrix $C$ to be diagonalized
at each iteration. Therefore, the entanglement entropy is computed in practice
as $-\sum_i\lambda_i^4\ln\lambda_i^4$ where the $\lambda_i$'s are proportional
to the eigenvalues of $C$ with the constraint $\sum_i\lambda_i^4=1$. In the
thermodynamic limit, the cut is the same as the one performed in the transfer
matrix approach so $S_A$ measures the entanglement between the left and right
halves of the systems or, more precisely, between the left and right
halves of the equivalent quantum spin chain.}
\\

In this work, the behavior of entanglement entropies are studied for a
model of non-overlapping $k$-mers deposited on a lattice. The case $k=2$
corresponds to the celebrated dimer model that has attracted a lot of
interest in the last half-century. Besides its experimental relevance to
systems where diatomic molecules are adsorbed on a surface~\cite{Fowler}, the
full covering of a graph by dimers was mostly studied by physicists and
mathematicians from a purely theoretical perspective~\cite{Kenyon,Kenyon2}.
Fisher~\cite{Fisher} and Kasteleyn~\cite{Kasteleyn} independently managed to
express the partition function as a pfaffian and then compute exactly the free
energy density of the model. The more general case of a mixture of monomers
and dimers was subsequently studied and, thanks to a mapping onto an Ising
model, it was shown that the free energy density is an analytic function of
the chemical potential of the dimers~\cite{Heilman,Heilman2}. Therefore, the
model does not undergo any phase transition. The same conclusion was drawn
for a lattice model of trimers~\cite{Dhar}. As recently shown, a
Kosterlitz-Thouless phase can nevertheless be observed in the dimer model at
close-packing when an interaction is introduced between aligned dimers on
the same plaquette of the square lattice~\cite{Alet1,Alet2}.
\\

On the other hand, a gas of infinitely long rigid polymers is expected to
undergo an entropy-driven first-order transition between an isotropic and a
nematic phase~\cite{Onsager,Frenkel2}. A discretization of the orientation
of the polymers does not change this conclusion~\cite{Zwanzig}. In 2D, a
generalization of the Mermin-Wagner theorem forbids the existence of a nematic
phase that would break the symmetry under rotation~\cite{MerminWagner,Straley}.
Monte Carlo simulations of infinitely thin needles have however shown
the existence of a Berezinskii-Kosterlitz-Thouless transition~\cite{Frenkel}.
For discrete orientations of the needles, the Mermin-Wagner theorem does not
hold anymore and a nematic phase may be observed.
\\

One may therefore assume that rigid finite polymers, consisting in $k$
monomers aligned on the lattice, should display an isotropic-nematic phase
transition for sufficiently large enough $k$. In 2007, Ghosh {\sl et al.} argued
that such a model should actually undergo two phase transitions as the
chemical potential is increased~\cite{Ghosh}. Like infinitely long rigid
polymers, $k$-mers are first expected to undergo a transition between an
isotropic and a nematic phase. When approaching close-packing at high chemical
potential, the system is expected to return to an isotropic phase.
Using Monte Carlo simulations, Ghosh et al. showed that this scenario is
indeed observed for $k\ge 7$. However, such Monte Carlo simulations based on
local removal/deposition of a single $k$-mer are very difficult due to a large
autocorrelation time. Nevertheless, the first isotropic-nematic transition
was shown to be continuous with critical exponents compatible with the Ising
universality class on the square lattice and the three-state Potts model
one on the triangular lattice~\cite{Matoz1,Matoz2,Matoz3,Linares}. This
critical behavior is explained by the fact that, in the nematic phase,
the $\mathbb{Z}_q$ symmetry of the different orientations of the $k$-mers
is spontaneously broken~\cite{Vink}. There are $q=2$ possible orientations
on the square lattice and $q=3$ on the triangular lattice. Later a cluster
algorithm updating $k$-mers along a whole row or column of the lattice
was introduced~\cite{Kundu1} and allowed for studying the model at high
densities. The second nematic-isotropic transition was shown to be
continuous too but the estimated critical exponents are incompatible
with the Ising universality class. The possibility of a cross-over to
Ising universality class at large length scales is
however not excluded by the authors.
\\

In this study, the $k$-mer model is considered on the square lattice.
A new Corner Transfer Matrix Renormalization Group (CTMRG) is introduced
to cope with the fact that the Corner Transfer Matrix of the $k$-mer
model is not symmetric for $k>2$, in contrast to usual lattice spin models.
The details of the algorithm are presented in the first section along with
the Boundary Conditions chosen to break the symmetry and the different
observables estimated to characterize the phase transitions. In particular, the
entanglement entropies are introduced. Results for the Ising, Potts and
clock models, undergoing respectively continuous, discontinuous and two
Berezinskii-Kosterlitz-Thouless transitions, are discussed to allow further
comparisons with the 7-mer model. In the second section, numerical
data for the 6, 7 and 8-mer models are presented and discussed.
In agreement with previous Monte Carlo simulations, the order parameter
and the entropy reveals the existence of a nematic phase for $k\ge 7$.
In the third section, numerical evidence is given that the entanglement
entropies of $k$-mer model does not diverge at the transitions.
Conclusions follow.

\section{Numerical methodology}
In 1968, Baxter introduced the first Matrix-Product-State algorithm for the
monomer-dimer model on the square lattice~\cite{Baxter2}. As in DMRG to be
introduced 25 years later~\cite{White1,White2,Schollwock1,Schollwock2},
the ground state of the
classical transfer matrix is approximated by a Matrix-Product-State (MPS).
The optimization of this MPS is performed by alternating between the transfer
matrices generating the lattice horizontally and vertically respectively. The
convergence to a machine-precision accuracy is extremely fast, mainly due to
the fact that the model is not critical. Two of the tensors forming the MPS
turn out to be corner transfer matrices. Shortly after the introduction of DMRG,
an algorithm, based on this corner transfer matrix and known as Corner Transfer Matrix
Renormalization-Group algorithm (CTMRG), was introduced for classical
systems~\cite{Nishino}. Neither the Baxter algorithm nor CTMRG can be applied
to $k$-mers with $k>2$ because the corner transfer matrix is not symmetric
in this case. In this paper, a new CTMRG is introduced for the $k$-mer model.
The symmetry of the corner-transfer matrix is not required anymore. The
algorithm exploits the mirror symmetry under reflection with respect to the
horizontal or vertical axis to greatly improve the convergence.

\subsection{Corner Transfer Matrix Renormalization Group algorithm}
\begin{figure}[ht]
\begin{center}
    \includegraphics[width=8cm]{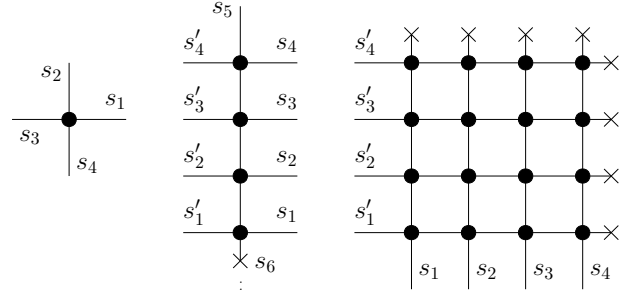}
	\caption{\CC Diagrammatic representation of the vertex {\AG $w$} (left),
    the transfer matrix {\AG $T$} (center) and the corner transfer matrix
    {\AG $C$} (right).
    Each black circle denotes a weight $w(s_1,s_2,s_3,s_4)$. The crosses
    means that some specific boundary conditions are applied: the state of
    the bond can be fixed or a sum can be performed over all possible values.
    The sum over the states of all the internal bonds is implicit.}
\label{fig0}
\end{center}
\end{figure}

{\CC Each vertex of the square lattice is given a statistical weight $w(s_1,
s_2,s_3,s_4)$ which depends on the states $s_1$, $s_2$, $s_3$ and $s_4$ of the four
incoming bonds. The elements of the {\AG transfer matrix $T$} are defined as the
statistical weight of a single row (or column) of vertices. They can
be written as a product of $w$'s. Using the notations of figure~\ref{fig0},
these elements reads
   \ba &&T^{s_5}(s_1,s_2,\ldots;s_1',s_2',\ldots)\nonumber\\
   &&=\prod_{s_6,s_7,\ldots}w(s_1,s_7,s_1',s_6)
   w(s_2,s_8,s_2',s_7)\ldots \ea
Note that the transfer matrix still depends on the state $s_5$
of the pending bond at the top. {\AG Depending on the specific type of
boundary conditions we intend to impose, the appropriate setting of the state
$s_6$ needs to be taken.}
The elements of the {\AG corner transfer matrix $C(s_1,s_2,\ldots;$
$s_1',s_2',\ldots)$} are the statistical weights of a square
(see figure~\ref{fig0})~\cite{Baxter}. Boundary Conditions have
been applied to the bonds on {\AG the two opposite} sides of the square.
}

\begin{figure*}[ht]
\begin{center}
    \includegraphics[width=14cm]{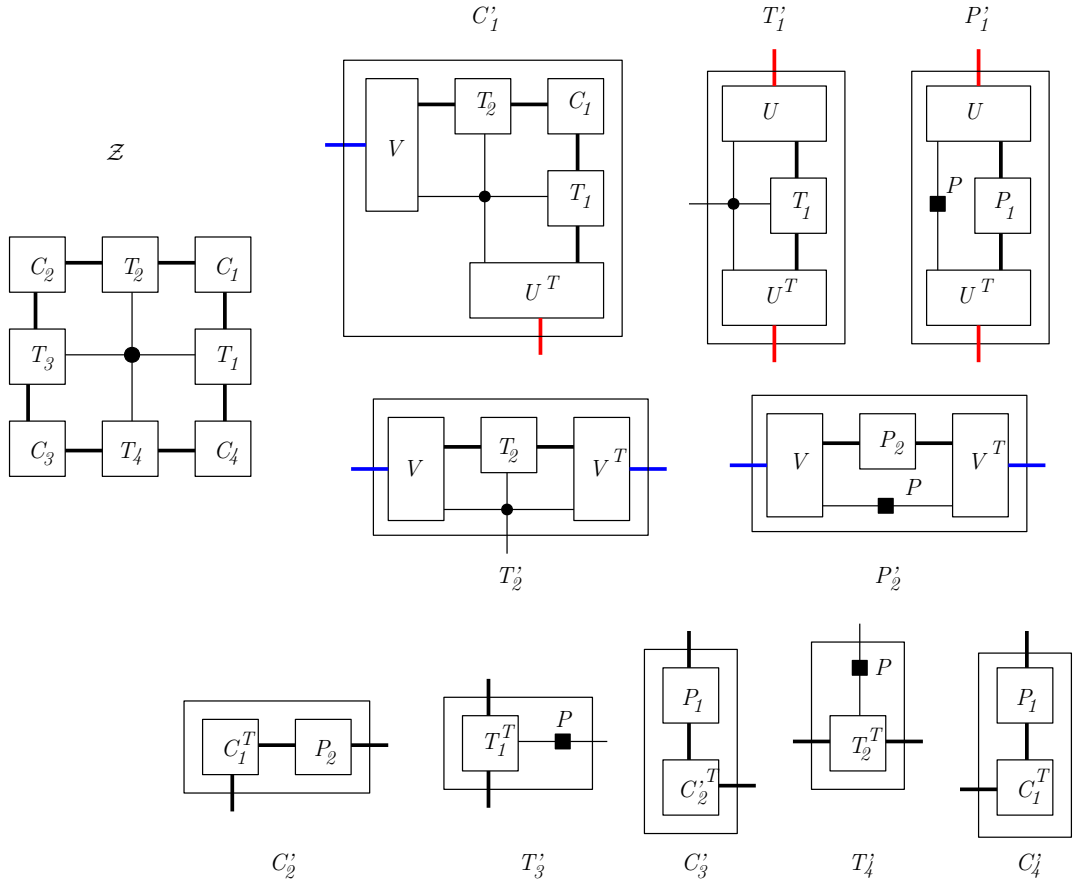}
    \caption{Diagrammatic description of the CTMRG algorithm for the $k$-mer
      model. On the left, the partition function $Z$ is decomposed into 8
      tensors, 4 transfer matrices $T_i$ and 4 corner transfer matrices $C_i$.
      The 4-leg black circle represents the statistical weight $w$ on the
      central vertex. On each internal line, a sum over all possible states
      is implicitly performed. Thick lines are associated to renormalized
      states while thin lines were not renormalized yet. In the general case,
      the tensors $T_i$ and $C_i$ are independent. At the center of the first
      line, the corner transfer matrix $C_1$ is first extended by contraction
      with the two transfer matrices $T_1$ and $T_2$ and a weight $w$. The
      resulting tensor is decomposed into singular values (SVD) corresponding
      to changes of basis $U$ and $V^T$ on its two external legs. Similarly,
      the transfer matrices $T_1$ and $T_2$ are extended by contraction with
      a weight $w$ and the same change of basis $U$ or $V$ is applied. Without
      any truncation of $U$ and $V$, the partition function is unchanged. On
      the right of the first and second lines, the horizontal and vertical
      mirror operators $P$ are extended and the change of basis is applied.
      Using these mirror symmetries, the corner transfer matrices $C_2$,
      $C_3$ and $C_4$ can be constructed from $C_1$. Transfer matrices $T_3$
      and $T_4$ are obtained from $T_1$ and $T_2$.}
  \label{fig1}
  \end{center}
\end{figure*}

The partition function can be decomposed into one vertex, four transfer
matrices $T_i$ and four corner transfer matrices $C_i$ as (figure~\ref{fig1})
  \be{\cal Z}=\sum_{s_1,s_2,s_3,s_4} w(s_1,s_2,s_3,s_4)
  \trace \big[T_1^{s_1}C_1T_2^{s_2}C_2T_3^{s_3}C_3T_4^{s_4}C_4\big].
  \label{ExprZ}\ee
The four parameters $s_1$, $s_2$, $s_3$, and $s_4$ of the vertex
correspond to the states of the bonds at the right, top, left and bottom
of the vertex.
This decomposition is diagrammatically represented on figure~\ref{fig1}.
The thin lines on the figure correspond to bonds that carry a single degree
of freedom. The thick lines carry renormalized states. In the following, we
are interested in systems for which the statistical weight of a vertex is
symmetric up to a local operation under a mirror transformation with
respect to both the vertical and horizontal axis:
  \ba
    &&w(s_1,s_2,s_3,s_4)=\sum_{s_1',s_3'}P_h(s_1,s_1')P_h(s_3,s_3')
    w(s_3',s_2,s_1',s_4),\nonumber\\
    &&w(s_1,s_2,s_3,s_4)=\sum_{s_2',s_4'}P_v(s_2,s_2')P_v(s_4,s_4')
    w(s_1,s_4',s_3,s_2')\nonumber\\
	\ea
with
   \be P_h=P_h^+,\hskip 1cm P_h^2=1,\hskip 1cm
   P_v=P_v^+,\hskip 1cm P_v^2=1 \ee
For spin models, the matrices $P_h$ and $P_v$ are equal to the identity. For
$k$-mer models, it will not be the case anymore. The transfer matrices $T_i$
and the corner transfer matrices $C_i$ are also expected to be symmetric under
these two mirror transformations. As a consequence, it is sufficient to consider
$C_1$ and the two transfer matrices $T_1$ and $T_2$. Other matrices will be
reconstructed from these three. The first step of the algorithm consists in
extending the corner transfer matrix by adding the two transfer matrices,
$T_1$ and $T_2$, and a vertex $w$:
    \ba C_1'((s_4,s_5),(s_3,s_8))=&&\sum_{s_1,s_2,s_6,s_7}w(s_1,s_2,s_3,s_4)\\
    &&\times T_1^{s_1}(s_5,s_6)C_1(s_6,s_7)T_2^{s_2}(s_7,s_8)\nonumber
    \ea
$(s_4,s_5)$ denotes a product state constructed from the states $s_4$ and $s_5$.
The process is represented diagrammatically on the figure (second diagram
from the left on the top row). To reduce the dimension of $C_1$, a decomposition
into singular values (SVD) is performed on $C_1$. The latter is replaced by
a diagonal matrix whose elements are the largest singular values $\Lambda$:
    \ba C_1''(s_9,s_{10})=&&U^T(s_9,(s_4,s_5))C_1'((s_4,s_5),(s_3,s_8))
    \nonumber\\
	&&\times V((s_3,s_8),s_{10})=\Lambda_{s_9}\delta_{s_9,s_{10}}.\ea
The number of singular values that are kept, and therefore the dimension of
$C_1''$, is a fixed parameter. The transfer matrices $T_1$ and $T_2$ are then
extended by contraction with a single vertex:
   \ba
    &&{T_1'}^{s_3}((s_4,s_5),(s_2,s_6))=\sum_{s_1}
    w(s_1,s_2,s_3,s_4)T_1^{s_1}(s_5,s_6),\nonumber\\
    &&{T_2'}^{s_4}((s_1,s_5),(s_3,s_6))=\sum_{s_2}
    w(s_1,s_2,s_3,s_4)T_2^{s_2}(s_5,s_6),\nonumber\\
    \ea
and then renormalized by performing the appropriate basis change:
   \be
    {T_1''}^{s_3}=U^T{T_1'}^{s_3}U,\hskip 1truecm
    {T_2''}^{s_4}=V{T_2'}^{s_4}V^T
    \ee
To construct the other transfer matrices, the mirror transformations need to
be extended and renormalized too. Setting initially $P_1=P_v$ and $P_2=P_h$,
the matrices are extended as
   \ba
   &&P_1'((s_1,s_2),(s_3,s_4))=P_v(s_1,s_3)P_1(s_2,s_4),\nonumber\\
   &&P_2'((s_1,s_2),(s_3,s_4))=P_h(s_1,s_3)P_2(s_2,s_4)
   \ea
i.e. $P_1'=P_h\otimes P_1$ and $P_2'=P_h\otimes P_2$ and then renormalized as
  \be P_1''=U^TP_1'U,\hskip 1cm P_2''=VP_2'V^T\ee
Finally, the other transfer matrices are given by
  \be C_2'=P_2C_1^T,\quad
  C_3'=P_1{C_2'}^T,\quad
  C_4'=C_1^TP_1=P_2{C_3'}^T\label{SymetryC}\ee

The algorithm can be iterated either until the lattice size reaches the
desired one or until thermodynamic averages become stable, i.e. lattice size
independent, up to some accuracy.
The number of iterations needed for convergence depends strongly on the
proximity of a phase transition and on the number of states kept during
the renormalization process. In the case of the $k$-mer model, we observed
that convergence depends also on the boundary conditions. A faster convergence
is usually obtained with random initial tensors. However, thermodynamic
averages display oscillations as the lattice size is increased and for
many points of the phase diagram, mostly in the nematic phase, we were
not able to reach convergence at large number of states. Therefore, in the
following, the study is limited to finite-size systems.

\subsection{Statistical weight of a vertex}
In the monomer-dimer model, each site of the lattice is occupied by a
monomer. Dimers correspond to a bond joining two neighboring sites. A
monomer can belong at most to one dimer. A configuration of the system is
therefore given by the set of bonds on which lies a dimer. A state, 0 or 1,
is assigned to all bonds of the lattice. $0$ indicates the absence of a
dimer while $1$ corresponds to the presence of a dimer. On a given site,
the statistical weight of an isolated monomer is
		\be w(0,0,0,0)=1,\ee
while for a monomer that belongs to an horizontal dimer
    \be w(1,0,0,0)=w(0,0,1,0)=e^{\mu_h}\ee
and to a vertical dimer
    \be w(0,1,0,0)=w(0,0,0,1)=e^{\mu_v}\ee
All other elements of $w$ are zero. Note that the factor $\beta=1/k_BT$ has
been absorbed into the definition of the chemical potentials $\mu_h$ and
$\mu_v$.
\smallskip

A $k$-mer correspond to a sequence of $k$ aligned monomers on the lattice.
It can also be seen as a sequence of $k-1$ dimers on $k-1$ successive bonds
of the lattice. The different dimers forming a $k$-mer needs to be
distinguished. Therefore, a bond can be in $k$ possible states. The state
0 denotes the absence of any dimer on the bond. Therefore, a vertex with
four bonds in the state 0 signals the presence of an isolated monomer.
The associated statistical weight is
    \be w(0,0,0,0)=1,\ee
The $k$-mers are decomposed into $k-1$ dimers labeled $1$ to $k-1$ from
left to right and bottom to top. For an horizontal $k$-mer, the statistical
weights are
   \ba&&w(1,0,0,0)=w(2,0,1,0)=\ldots\nonumber\\
   &&=w(k-1,0,k-2,0)=w(0,0,k-1,0)=e^{\mu_h}\ea
while, for a vertical dimer, the non-vanishing elements are
   \ba&&w(0,1,0,0)=w(0,2,0,1)=\ldots\nonumber\\
   &&=w(0,k-1,0,k-2)=w(0,0,0,k-1)=e^{\mu_v}.\ea
The image of a $k$-mer under a mirror transformation is still a $k$-mer but
with dimers labeled in the reversed order. Therefore, the mirror tensors
$P_h$ and $P_v$ satisfies
   \be P_{h,v}(k-1,1)=P_{h,v}(k-2,2)=\ldots=1\ee
and all other elements vanish.
\smallskip

To be able to compare the entropy and the entanglement entropy of $k$-mer
model with those of well-known lattice spin models, the algorithm was also
applied to the $q$-state Potts and clock models. In both cases, a $q$-state
spin degree of freedom is placed on each bond of the square lattice. The
vertex considered above is therefore a plaquette of four spins. The
statistical weight is
   \be w(s_1,s_2,s_3,s_4)=e^{\big[V(s_1,s_2)
   +V(s_2,s_3)+V(s_3,s_4)+V(s_4,s_1)\big]/k_BT}\ee
where $V(s,s')=\delta_{s,s'}$ for the Potts model and $V(s,s')
=\cos{2\pi\over q}(s-s')$ for the $q$-state clock model.

\subsection{Boundary conditions}
Different boundary conditions can be imposed to the system, provided that
they are symmetric under horizontal and vertical mirror transformations.
For Open Boundary Conditions (OBC), the initial $T_1$ and $C_1$ tensors
are
    \ba &&{T_1}^{s_3}(s_4,s_2)=\sum_{s_1}w(s_1,s_2,s_3,s_4),\nonumber\\
    &&C_1(s_4,s_3)=\sum_{s_1,s_2} w(s_1,s_2,s_3,s_4)\ea
For Fixed Boundary Conditions (FBC) in the state $s=1$ for example,
they are chosen to be
    \be {T_1}^{s_3}(s_4,s_2)=w(1,s_2,s_3,s_4),\quad
    C_1(s_4,s_3)=w(1,1,s_3,s_4).\ee
In the $k$-mers model with $k\ge 7$, the system is expected to be in a
nematic phase for intermediate chemical potentials. In the latter, the
$k$-mers are mostly either horizontal or vertical ($\mathbb{Z}_2$
symmetry). In the second case, horizontal bonds are in the state 0
while vertical ones are in states between $1$ and $k$. To break
the $\mathbb{Z}_2$ symmetry of the nematic phase, mixed boundary
conditions are imposed on the system. On the left and right boundaries,
the horizontal bonds are forced to be in the state 0. On the upper
and lower boundaries, vertical bonds can be in any of the states
$1$ to $k-1$ but not $0$. This conditions are implemented in the initial
tensors as
    \ba C_1(s_4,s_3)&=&\sum_{s_2=1}^{k-1} w(0,s_2,s_3,s_4),\nonumber\\
    {T_1}^{s_3}(s_4,s_2)&=&w(0,s_2,s_3,s_4),\nonumber\\
	{T_2}^{s_4}(s_1,s_3)&=&\sum_{s_2=1}^{k-1} w(s_1,s_2,s_3,s_4)\ea
Other initial tensors are obtained by applying the mirror transformations
(\ref{SymetryC}).

\subsection{Observables}
\begin{figure}
  \begin{center}
    \psfrag{beta}[Bl][Bl][1][1]{$\beta$}
    \psfrag{S}[Bl][Bl][1][1]{$S$}
    \psfrag{Se}[Bl][Bl][1][1]{$S_A$}
    \includegraphics[width=8cm]{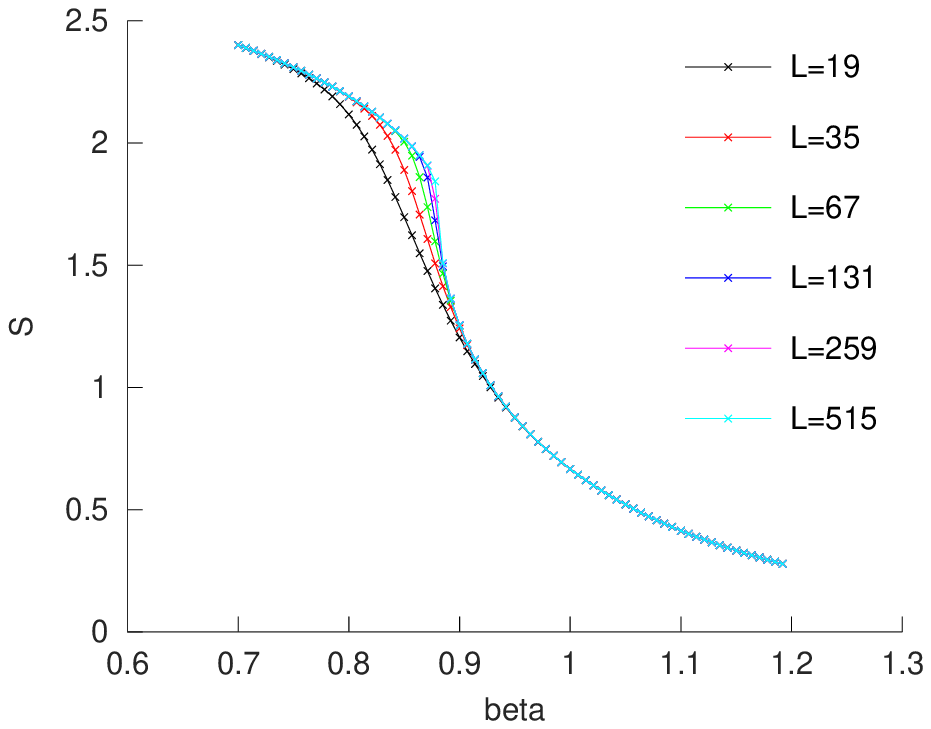}\quad
	\includegraphics[width=8cm]{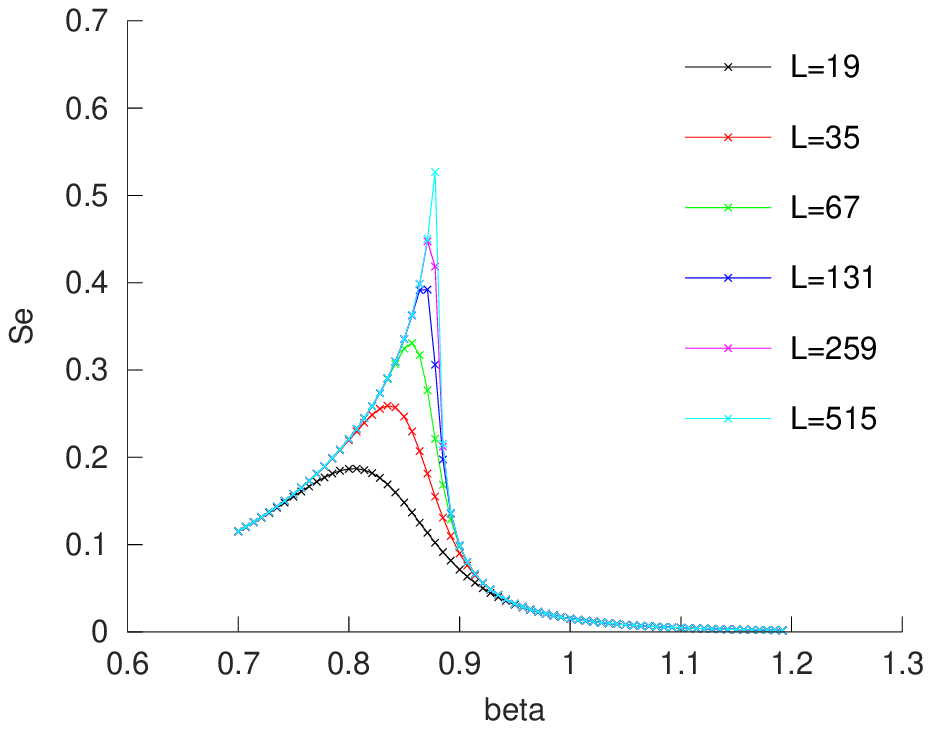}
	\caption{Entropy at the central vertex (top) and von Neumann entanglement entropy (bottom)
	of the 2-state Potts model (equivalent to the Ising model)
	with Fixed Boundary Conditions versus the inverse of the temperature $\beta=1/k_BT$.
	The different curves correspond to the different lattice sizes indicated
	in the legend. The data were computed with 32 states.}
  \label{fig2}
  \end{center}
\end{figure}

\begin{figure}
  \begin{center}
    \psfrag{beta}[Bl][Bl][1][1]{$\beta$}
    \psfrag{S}[Bl][Bl][1][1]{$S$}
    \psfrag{Se}[Bl][Bl][1][1]{$S_A$}
    \includegraphics[width=8cm]{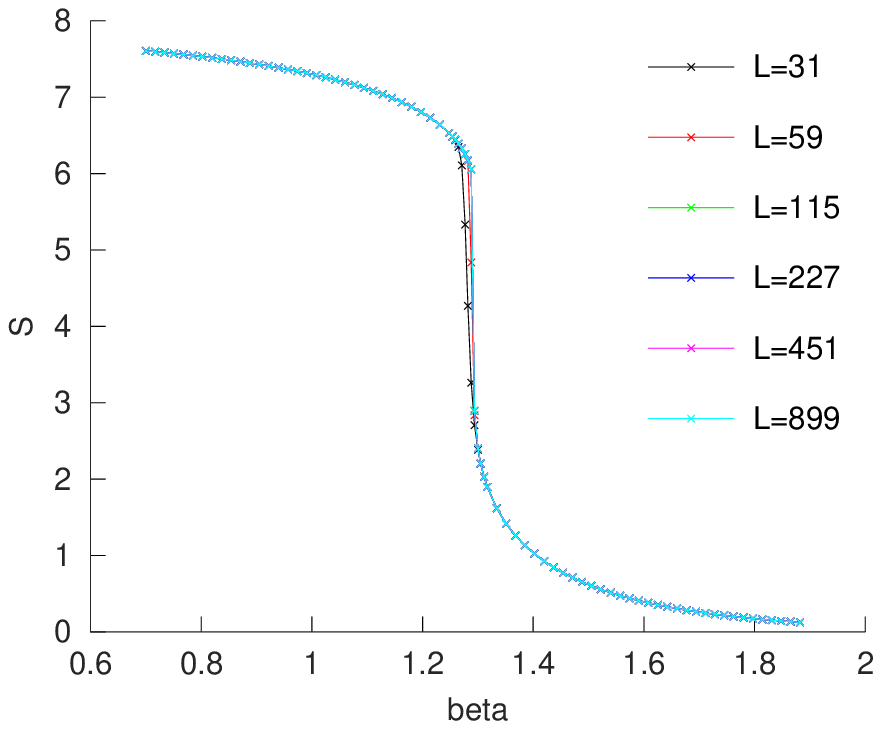}\quad
	\includegraphics[width=8cm]{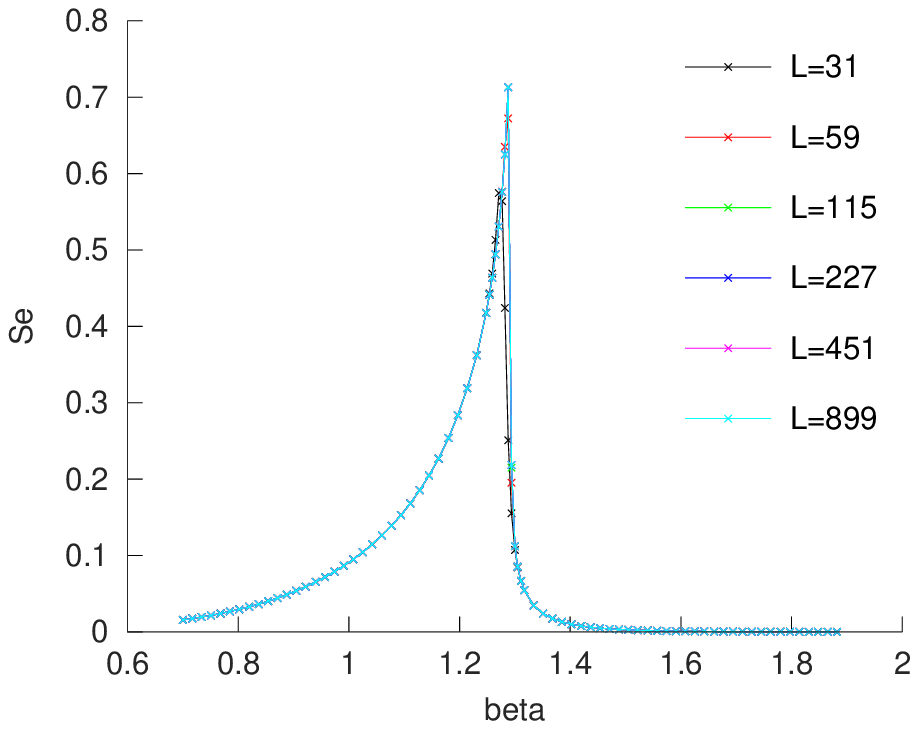}
	\caption{Entropy of the central vertex (top) and von Neumann entanglement entropy (bottom)
	of the 7-state Potts model with Fixed Boundary Conditions
	versus the inverse of the temperature $\beta=1/k_BT$.
	The different curves correspond to the different lattice sizes indicated
	in the legend. The data were computed with 147 states.}
  \label{fig3}
  \end{center}
\end{figure}

\begin{figure}
  \begin{center}
    \psfrag{beta}[Bl][Bl][1][1]{$\beta$}
    \psfrag{S}[Bl][Bl][1][1]{$S$}
    \psfrag{Se}[Bl][Bl][1][1]{$S_A$}
    \includegraphics[width=8cm]{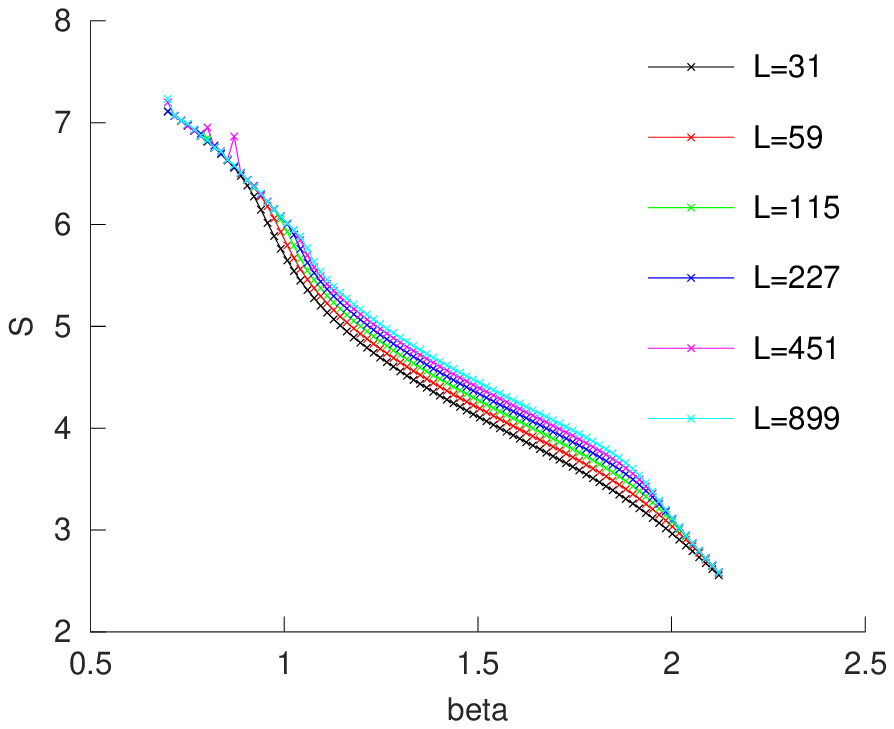}\quad
	\includegraphics[width=8cm]{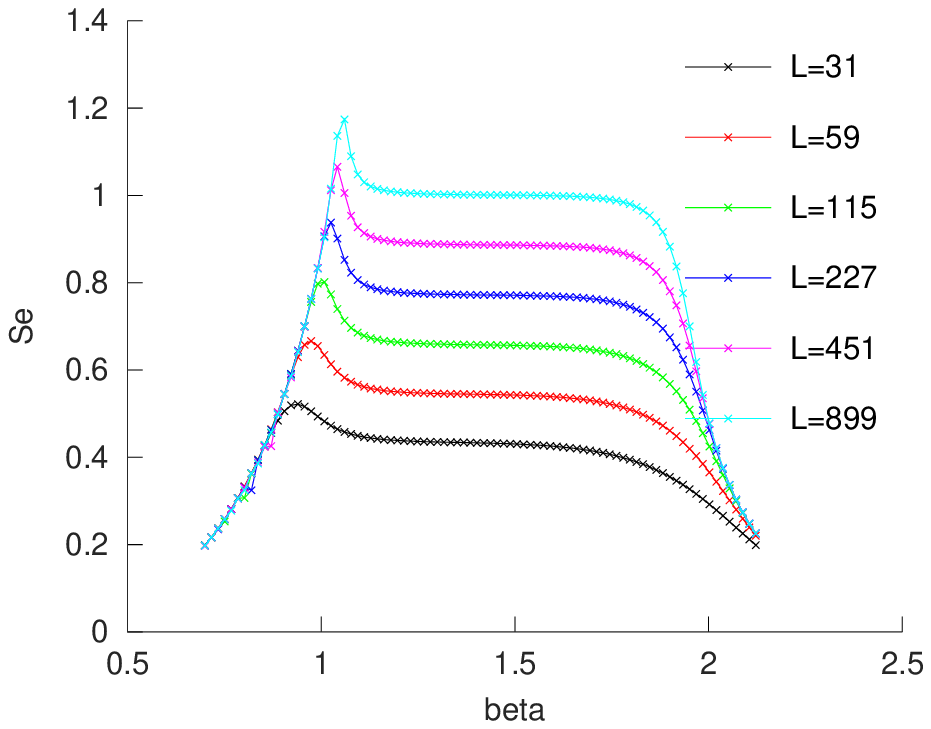}
	\caption{Entropy of the central vertex (top) and von Neumann entanglement entropy (bottom)
	of the 7-state clock model with Fixed Boundary Conditions
	versus the inverse of the temperature $\beta=1/k_BT$.
	The different curves correspond to the different lattice sizes indicated
	in the legend. The data were computed with 343 states.}
  \label{fig4}
  \end{center}
\end{figure}

The free energy density $f$ can be estimated from the partition function
${\cal Z}$. However, the convergence of this estimator is quite slow.
A much faster convergence is obtained with the estimator
     \be f=-\log\max T_1.\ee
Finite differences of this free energy at two close chemical potentials
$\mu$ and $\mu+\Delta\mu$ give access to the average density of $k$-mers
     \be\langle n\rangle=-{\partial f\over\partial\mu}
     \simeq -{f(\mu+\Delta\mu)-f(\mu)\over\Delta\mu}.\ee
This estimator is quite stable with $\Delta\mu\simeq 10^{-2}$. It is
nevertheless more convenient to measure the average density on the
central vertex. The statistical weight on the central vertex is indeed
easily computed as
    \ba\rho(s_1,s_2,s_3,s_4)=&&{1\over{\cal Z}}w(s_1,s_2,s_3,s_4)\nonumber\\
    &&\times\trace\big[T_1^{s_1}C_1T_2^{s_2}C_2T_3^{s_3}C_3T_4^{s_4}C_4\big]
    \label{StatCentral}\ea
{\CCb which corresponds to connecting the four arms of the central vertex
to four Transfer Matrices and inserting four Corner Transfer Matrices to
recover the square lattice. The construction is similar to (\ref{ExprZ})
for the partition function. At the $n$-th iteration of the algorithm,
the total lattice size is therefore $L=2n+3$.}
Average densities of horizontal (vertical) $k$-mers $\langle n_h\rangle$
($\langle n_v\rangle$) are computed on the two horizontal (vertical) bonds
of the central vertex as
    \be\langle n_{h,v}\rangle=\trace \rho n_{h,v}\label{avgDensity}\ee
where $n_{h,v}(s_1,s_2,s_3,s_4)=1$ when there is an horizontal (vertical)
dimer in the bond configuration $(s_1,s_2,s_3,s_4)$.
Due to the boundary conditions imposed to the system, these two densities
take different values in the nematic phase. An order parameter is then
defined as
     \be Q={\langle n_h\rangle-\langle n_v\rangle
     \over \langle n_h\rangle+\langle n_v\rangle}.\label{OrderParameter}\ee
The entropy of the central vertex embedded in the rest of the system
can also be computed as
   \be S=-\sum_{s_1,s_2,s_3,s_4}\rho(s_1,s_2,s_3,s_4)
   \log\rho(s_1,s_2,s_3,s_4)\label{Entropy}\ee
using the statistical weight (\ref{StatCentral}). It is a strictly local quantity
that should not be confused with the entropy per site. $S$ is nevertheless expected
to be singular at phase transitions. To allow for comparison and identify the
nature of the phase transitions of the 7-mer model, different spin models
undergoing second-order, first-order and topological phase transitions were
studied using the same CTMRG algorithm. As shown on figure~\ref{fig2}, a break
in the slope of $S$ can be observed at the critical temperature of the Ising
model. {\CC In the paramagnetic phase, the data shows that $S$ behaves as
$|\beta-\beta_c|\ln |\beta-\beta_c|$ over a broad range of temperatures.}
This is also the case in the paramagnetic phase with Open Boundary Conditions.
For the 7-state Potts model, which undergoes a first-order phase transition,
the entropy $S$ displays a jump at the transition temperature
(figure~\ref{fig3}). In the clock model, the entropy is observed to increase
with the lattice size in the intermediate critical phase (figure~\ref{fig4}).

\subsection{Entanglement entropies}
{\CC As mentioned in the introduction, {\AG an} $L\times L$ square lattice is considered
and a cut of size $L/2$ is made from the center of the square to the midpoint
of one of its edges (see figure~\ref{Se}). $L/2$ bonds are pending above the
cut and $L/2$ below. Consider {\AG a partially summed} partition function
{\AG ${\tilde{\cal Z}}_{s_1,s_2,\ldots}^{s_1',s_2',\ldots}$} of the system as a function
of the {\AG not summed-up} states $s_1,s_2,\ldots$ above the cut and $s_1',s_2',\ldots$
below the cut. {\AG The partition function ${\cal Z}$ can be reconstructed in the following}
{\AG
   \be{\cal Z}=\sum\limits_{s_1,s_2,\ldots,s_1',s_2',\ldots}
   {\tilde{\cal Z}}_{s_1,s_2,\ldots}^{s_1',s_2',\ldots}\ \
   \delta_{s_1,s_2,\ldots,s_1',s_2',\ldots}.\ee
} The quantity
{\AG
   \be\rho_A(s_1,s_2,\ldots;s_1',s_2',\ldots)=
   {{\tilde{\cal Z}}_{s_1,s_2,\ldots}^{s_1',s_2',\ldots}\over{\cal Z}}\label{partZ}\ee
} can be interpreted as the elements of a reduced density matrix. The definition
of the entanglement entropy between the degrees of freedom lying in the left
and righ halves of the system follows:
    \be S_A=-\trace\rho_A\ln\rho_A.\ee
By construction, the partition function of the square lattice with a cut
is given by the fourth power of the Corner Transfer Matrix~\cite{Peschel}:
     \be\rho_A={C^4\over\trace C^4}.\label{rdm}\ee
}

\begin{figure}[ht]
\begin{center}
    \includegraphics[width=5cm]{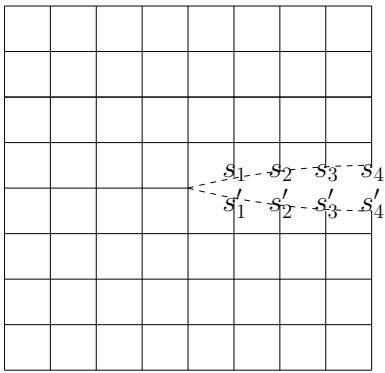}
	\caption{{\CC Cut made {\AG on the right side of} a square lattice
	{\AG defines the reduced density matrix $\rho_A$ in order} to measure the
	entanglement entropy between the degrees of freedom lying on the
	left and the right rectangular-shaped halves of the square lattice.}.
	}
\label{figSe}
\end{center}
\end{figure}

{\CC For spin models (Ising, Potts, clock, $\ldots$), the entanglement
entropy is easily computed from the eigenvalues of the Corner Transfer Matrix.}
{\AG Taking the partial sum of the four Corner Transfer Matrices,
cf. Eqs.~(\ref{partZ}) and (\ref{rdm}), results in the reduced density
matrix $\rho_A$ well-known in CTMRG~\cite{Nishino}}.
In the case of the $k$-mer model, the corner transfer matrices $C_i$
are not symmetric. Nevertheless, the relations (\ref{SymetryC}) show
that the reduced density matrix can be written as
    \be \rho_A=C_4C_3C_2C_1=(C_1^TC_1)^2.\ee
The matrix $C_1^TC_1$ is symmetric and its eigenvalues are the square
of the singular values of $C_1$ computed at each iteration. Therefore,
the von Neumann entanglement entropy is easily computed as
    \be S_A=-\sum_i \Lambda_i^4\ln\Lambda_i^4       \label{Se}\ee
where the $\Lambda_i$'s are proportional to the singular values of $C_1$
with the constraint $\sum_i\Lambda_i^4=1$. The Renyi entropy is defined
as
    \be S_n={1\over 1-n}\ln\big[\sum_i\Lambda_i^{4n}\big].\label{S2}\ee
In contrast to the entropy of the central vertex previously defined, the
entanglement entropy is a non-local quantity that depends on long-range
correlations in the system.
\\

Numerical data for the 2-state Potts model (equivalent to the Ising model)
are presented on the right of figure~\ref{fig2}. As the lattice size is
increased, the peak of the entanglement entropy becomes sharper and occurs
at a temperature closer to the critical point $\beta_c=\ln(1+\sqrt 2)
\simeq 0.881$. Numerical data for the entanglement entropy of the 7-state
Potts model, which undergoes a first-order phase transition at
$\beta_t=\ln(1+\sqrt 7)\simeq 1.294$, are presented on the right of
figure~\ref{fig3}. A sharp peak and a discontinuity are observed at the transition
temperature. Finally, the entanglement entropy of the 7-state clock
model is presented on figure~\ref{fig4}. As shown in~\cite{Krcmar2}, the
entanglement entropy grows with the lattice size in the whole intermediate
critical phase. The two Kosterlitz-Thouless phase transitions are revealed
by two peaks observed respectively with FBC and OBC.

\begin{figure}
  \begin{center}
    \psfrag{log(L)}[Bl][Bl][1][1]{$\ln L$}
    \psfrag{n}[Bl][Bl][1][1]{$n$}
    \psfrag{Se}[Bl][Bl][1][1]{$S_A$}
    \psfrag{f}[Bl][Bl][1][1]{$c_{\rm eff.}$}
    \includegraphics[width=8cm]{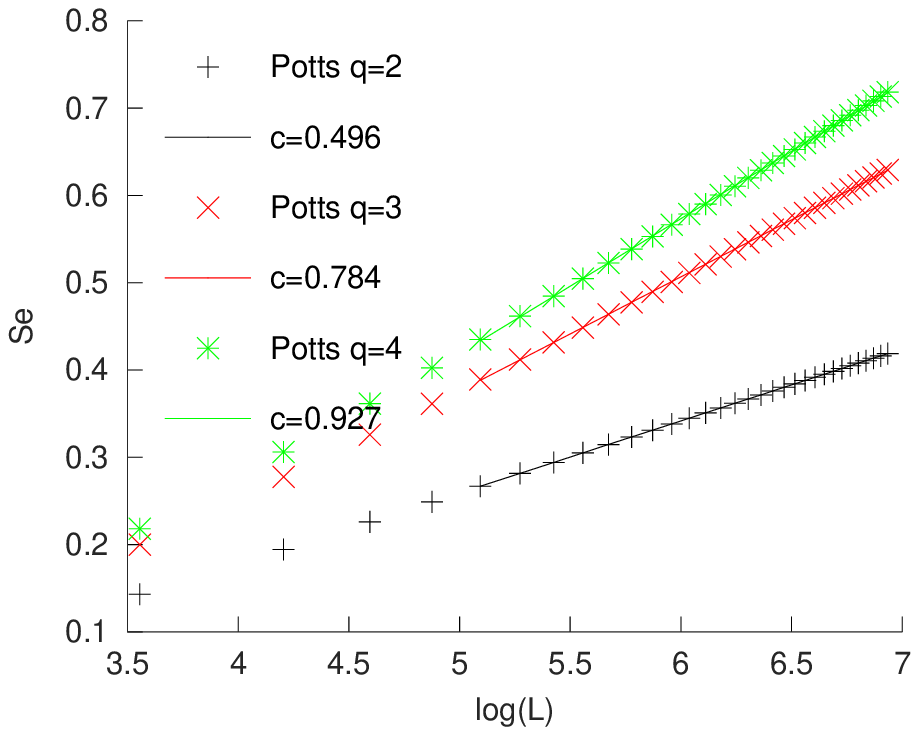}\quad
	\includegraphics[width=8cm]{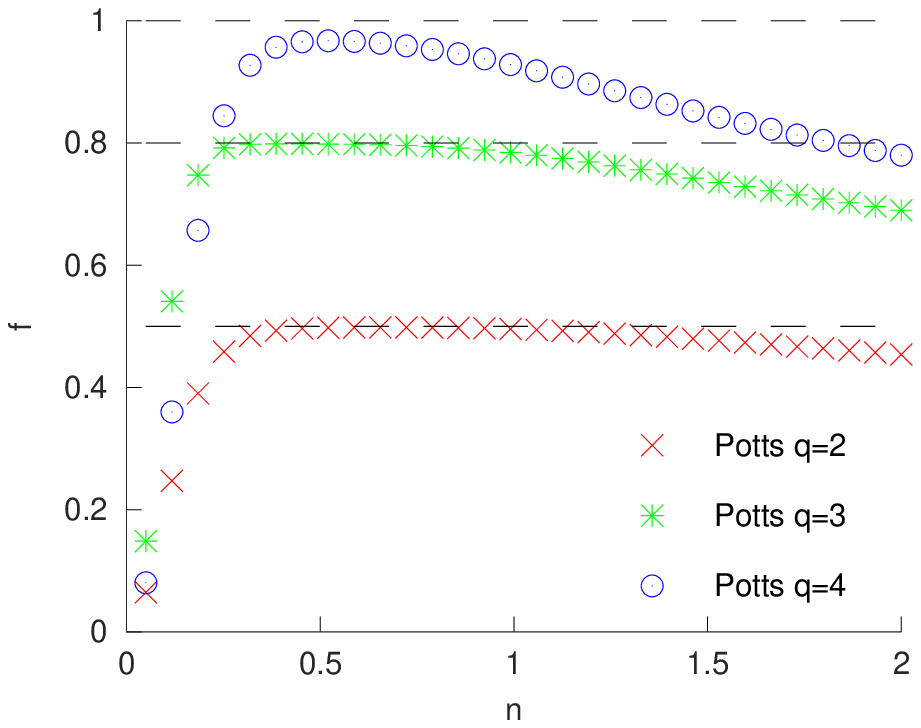}
	\caption{On the top figure, Finite-Size Scaling of the von Neumann entanglement
	entropies $S_A$ of the $q=2$, 3 and 4-state Potts models at their critical
	point $\beta_c=\log(1+\sqrt q)$. The continuous lines are linear fits of
	the data. The central charges, estimated from the slopes, are given in
	the legend. On the bottom figure, the effective charges extracted from a linear
	fit of the Renyi entropies $S_n$ are plotted versus $n$. The expected
	values are displayed as dashed lines.}
  \label{fig13}
  \end{center}
\end{figure}

As mentioned in the introduction, at the critical point the entanglement entropies
are expected to scale with the length $\ell$ of the cut as
    \be S_A\sim {c\over 6}\ln\ell,\quad\quad\quad
    S_n\sim {c\over 12}\left(1+{1\over n}\right)\ln\ell. \label{FSS_Se}\ee
On figure~\ref{fig13}, the entanglement entropies $S_A$ and $S_2$ are plotted
versus the logarithm of the size $L$ of the cut for the $q=2$, 3 and 4-state
Potts models at their critical point. As expected, the entropies display a
linear behavior with $\log L$. Assuming that, for the von Neumann entanglement
entropy $S_A$, the prefactor is the same as in the strip geometry (\ref{FSS_Se}),
the estimates of the central charge are compatible with the known values
$c=1$, $4/5$ for the 2 and 3-state Potts models respectively. The data for the
4-state Potts model, whose central charge is known to be $c=1$, are slightly
away from the theoretical prediction, maybe due to the fact that the critical
point is actually a tricritical point involving stronger corrections.
For the Renyi entropy $S_n$, the expected prefactor is recovered for values of
$n$ around $n\simeq 1/2$. For larger values of $n$, strong deviations are observed.

{\CC
\subsection{Convergence and error bars}
Provided that there is no source of systematic deviations (due to insufficient
thermalization or lack of ergodicity), error bars in Monte Carlo simulations
result only from statistical fluctuations. The latter can be made as small as
desired by increasing the number of Monte Carlo steps. In CTRMG, the only
source of error is the truncation of the Corner Transfer Matrix. The dimension
of the latter increases exponentially fast with the lattice size.
If all states were kept, the calculation would be exact. By increasing
the number of states kept at each truncation, the deviation from the exact
result can be made smaller. However, in contrast to Monte Carlo simulations,
it is not possible to estimate the systematic deviation introduced by the
truncation of the Corner Transfer Matrix. Therefore, in the rest of the
paper, the observables are usually plotted for different number of states
to show the convergence of the data.
\\

In DMRG studies, the level of approximation is often quantified with the so-called
truncation error. The latter is defined as the sum of the eigenvalues of the
density matrix that are discarded. The equivalent of the density matrix in CTMRG
is $C^4$, the fourth power of the Corner Transfer Matrix. In our implementation,
a Singular Value Decomposition of $C$ is performed at each iteration and a small
number $m$ of
singular values $\Lambda$ are kept while the rest is discarded and the matrices
are truncated. A possible definition of a truncation error for CTMRG is therefore
    $$\varepsilon={\sum_{i=m}^N \Lambda_i^4\over \sum_{i=1}^N\Lambda_i^4}$$
The smallest the truncation error and the more accrate the simulation. However,
it is not possible to establish a simple relation between the truncation error
and the error bar on the observables computed in CTMRG (free energy, order
parameter entanglement entropy, $\ldots$). It is a major drawback of the method,
shared with DMRG and all Matrix-Product and Tensor-Network algorithms.
In the simulations whose results are presented in the paper, only the $m$ largest
singular values of the Corner Transfer Matrix are computed at each iteration
using the {\tt Arpack} library. To discuss the behavior of the truncation error
with the simulation
parameters, we have implemented a (slower) version of the code where all singular
values are computed (with {\tt Lapack}). It allows for the computation of the
truncation error at each iteration but for smaller systems.
}

\begin{figure}
  \begin{center}
    \psfrag{m}[Bl][Bl][1][1]{$m$}
    \psfrag{mu=0}[Bl][Bl][1][1]{$\mu=0$}
    \psfrag{mu=0.6}[Bl][Bl][1][1]{$\mu=0.6$}
	\psfrag{mu=1.2}[Bl][Bl][1][1]{$\mu=1.2$}
	\psfrag{Err.}[Bl][Bl][1][1]{$\varepsilon$}
    \includegraphics[width=8cm]{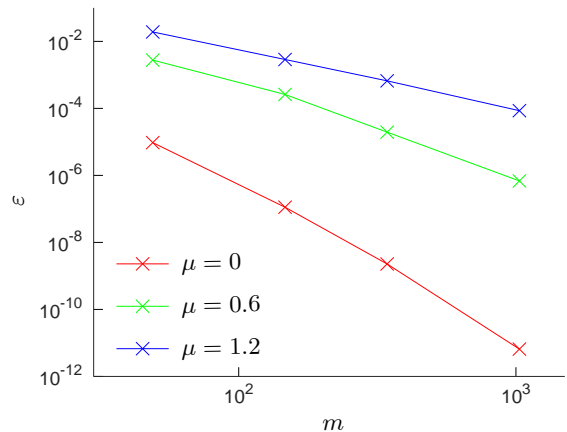}\
    \caption{{\CC Truncation error of the 7-mer model versus the number of states
    $m$ for $\mu=0,0.6$ and $1.2$ and a lattice size $L=51$.}}
   \label{figErr}
  \end{center}
\end{figure}

{\CC As can be seen on figure~\ref{figErr} for the 7-mer model, the truncation
error increases very rapidly with the chemical potential. The accuracy
is therefore expected to be much better in the low-density phase than in the
high-density phase. The truncation error displays a decay with the number of
states $m$ which is close to a power-law. For $\mu=1.2$, the exponent of this
power-law decay is $-1.8$.
}

\section{Numerical evidences of phase transitions in the $k$-mer model}
The 7-mer model is studied by means of the CTMRG algorithm keeping a number of states
equal to 343, 686 or 1029. During each simulation, the calculation is stopped after
14, 28, 56, 112, 224 and 448 iterations, which corresponds respectively to lattice
sizes 31, 59, 115, 227, 451 and 899. The different observables are then computed.
To allow for comparison, the 6-mer and 8-mer model were also studied. In the latter,
512 states were kept and the observables were computed for the lattice sizes 35, 67,
131, 259, 515 and 1027. For the former, 648 states were kept and the observables
were computed for the lattice sizes 27, 51, 99, 195, 387 and 771.

\subsection{Average density}
\begin{figure}
  \begin{center}
    \psfrag{mu}[Bl][Bl][1][1]{$\mu$}
    \psfrag{n}[Bl][Bl][1][1]{$\langle n\rangle$}
    \includegraphics[width=8cm]{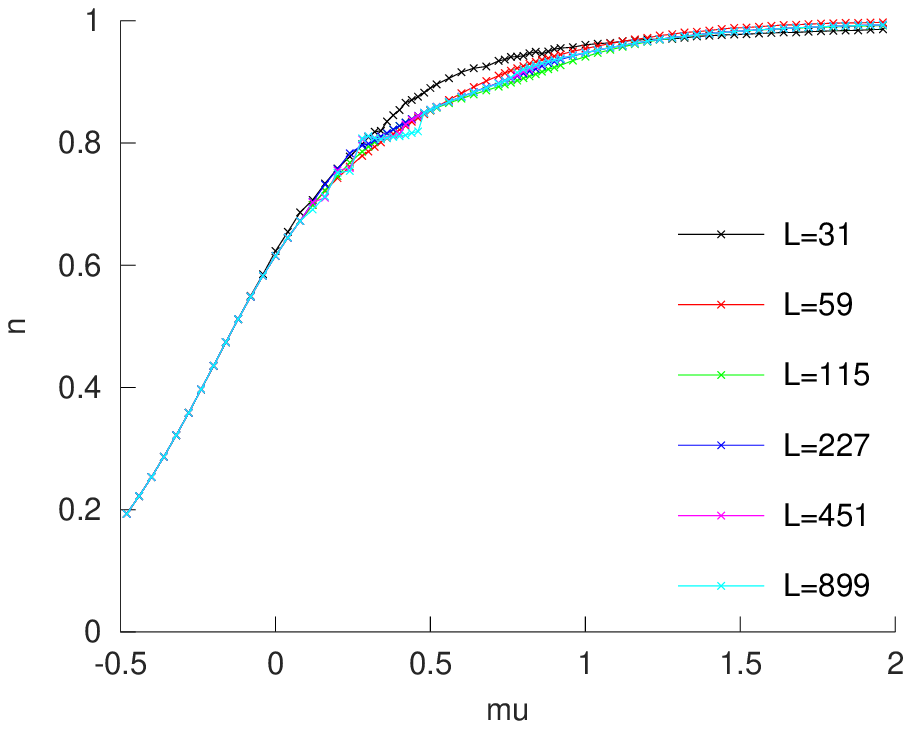}\quad
    \includegraphics[width=8cm]{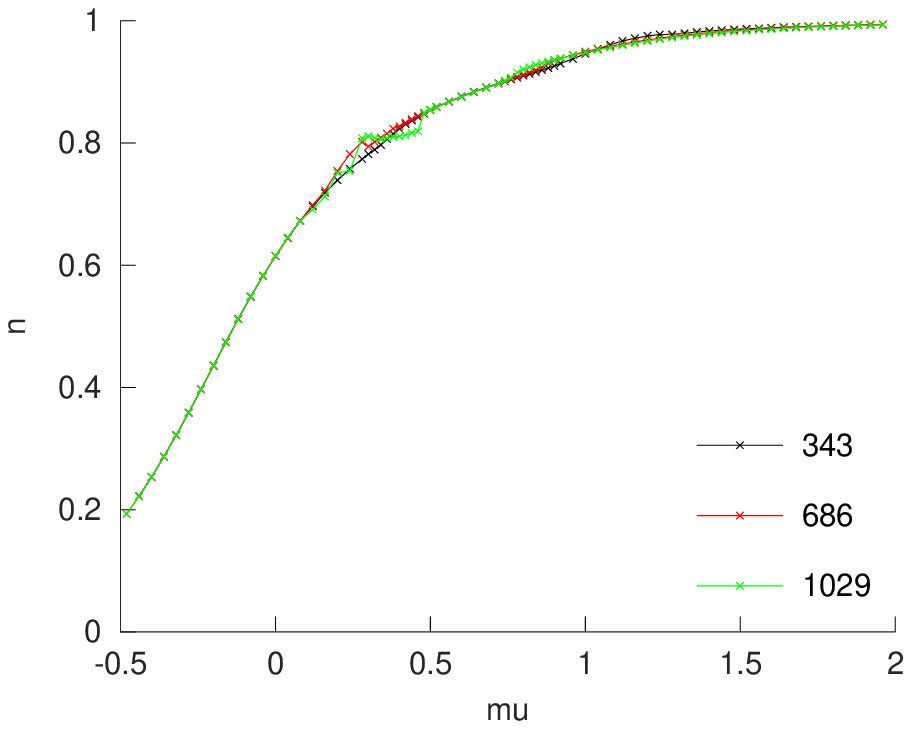}
    \caption{Average density of the 7-mer model model versus the chemical potential $\mu$
    per monomer. On the top figure, the data were computed using CTMRG with 1029 states
    and the different curves correspond to different lattice sizes $L$ as indicated by the legend.
    On the bottom figure, the lattice size is fixed to $L=899$ but different numbers of states were
    kept in the CTMRG algorithm (343 in black, 686 in red and 1029 in green).}
  \label{fig5}
  \end{center}
\end{figure}

As shown on figure~\ref{fig5}, the average total density $\langle n\rangle
=\langle n_v\rangle+\langle n_h\rangle$ (\ref{avgDensity}) of 7-mers increases monotonously
with the chemical potential $\mu$. For negative chemical potentials, the average density
depends only very weakly on the lattice size. In the intermediate range $0.2-1.0$,
stronger finite-size effects are observed for the smallest lattice size, $L=31$.
For large chemical potentials, finite-size corrections are again weaker but a different
sign than in the intermediate region. The assumption of the existence of three different
phases, as made in the literature on the basis of Monte Carlo simulations, would fit
with these observations. Note that the same observations can be made from the data
of the 8-mer model. For the 6-mer model, for which no transition is expected, finite-size
effects are nevertheless observed at intermediate chemical potentials.

\subsection{Order parameter of the nematic phase}
\begin{figure}
  \begin{center}
    \psfrag{mu}[Bl][Bl][1][1]{$\mu$}
    \psfrag{Q}[Bl][Bl][1][1]{$Q$}
    \includegraphics[width=8cm]{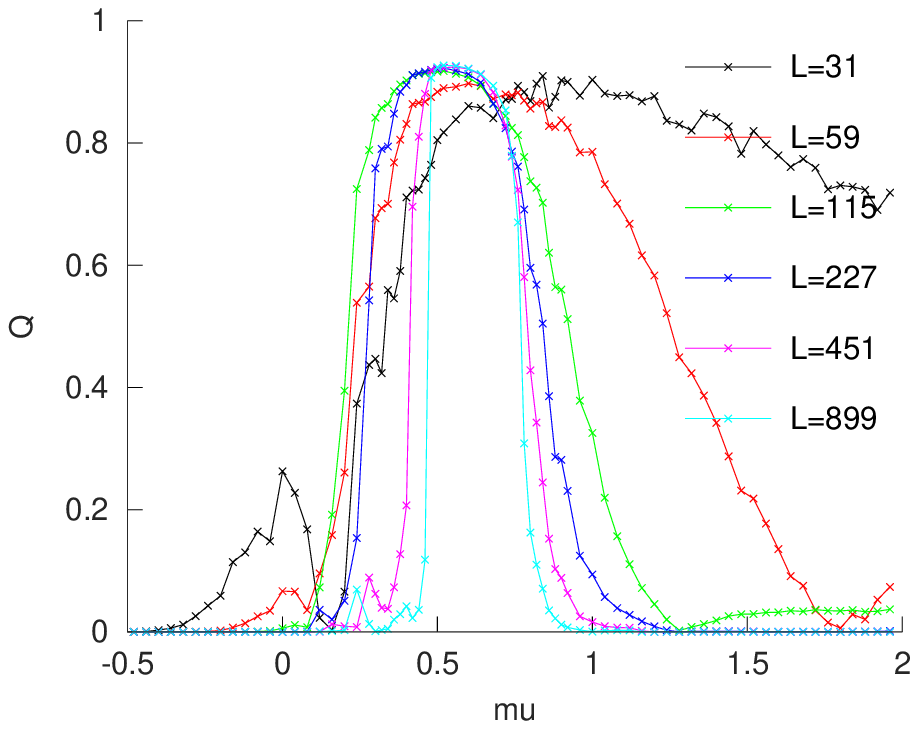}\quad
    \includegraphics[width=8cm]{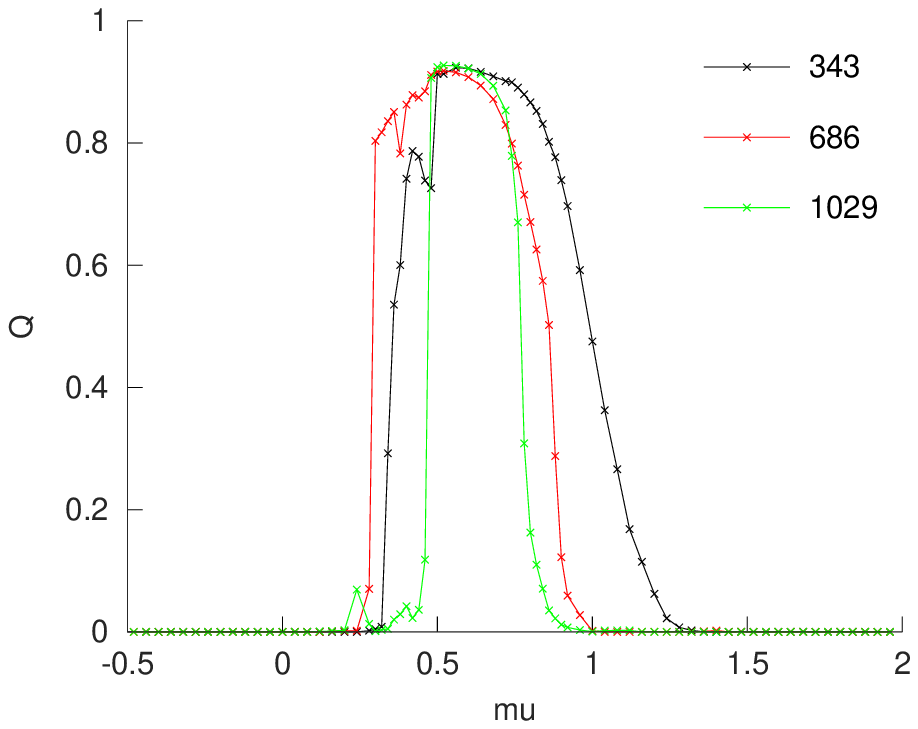}
    \caption{Order parameter of the nematic phase of the 7-mer model versus the chemical
    potential $\mu$ per monomer. On the top figure, the data were computed using CTMRG with 1029 states
    and the different curves correspond to different lattice sizes $L$ as indicated by the legend.
    On the bottom figure, the lattice size is fixed to $L=899$ but different numbers of states were
    kept in the CTMRG algorithm (343 in black, 686 in red and 1029 in green).
    }
  \label{fig6}
  \end{center}
\end{figure}

On figure~\ref{fig6}, the order parameter $Q$ (\ref{OrderParameter}) of the 7-mer model
is plotted versus the chemical potential. Figure~\ref{fig6} (right) shows that the location
of the phase boundaries depends non only on the lattice size but also on the number of states
kept during the CTMRG calculation. Despite the important computational effort devoted to
this study, an extrapolation of the chemical potentials at the transition remains elusive.
For sufficiently large lattice sizes, the same shape as observed in Monte Carlo
simulations~\cite{Ghosh} is obtained with CTMRG. For a lattice size $L=899$ and keeping
1029 states during the renormalization of the corner transfer matrix, the boundaries
of the nematic phase, signaled by a non-vanishing order parameter $Q$, can be estimated
at the chemical potentials per monomer $\mu_1\simeq 0.46$ and $\mu_2\simeq 0.91$. The second
value is quite far from the Monte Carlo estimate $\mu_2\simeq 0.795$~\cite{Kundu1}.
However, figure~\ref{fig6} shows that the nematic phase tends to shrink as the lattice
size is increased so an extrapolation may eventually give a closer estimate of $\mu_2$
in the thermodynamic limit. The average densities at the transitions are estimated
to be $\langle n_1\rangle\simeq 0.83$ and $\langle n_2\rangle\simeq 0.91$ to be compared
with the Monte Carlo estimates $\langle n_1\rangle\simeq 0.745$ and $\langle n_2\rangle
\simeq 0.915(15)$. In contrast to the chemical potentials, the CTMRG and Monte Carlo
estimates of the density at the second transition are nicely compatible due to the
fact that the average density varies slowly with the chemical potential.

\begin{figure}
  \begin{center}
    \psfrag{mu}[Bl][Bl][1][1]{$\mu$}
    \psfrag{Q}[Bl][Bl][1][1]{$Q$}
     \includegraphics[width=8cm]{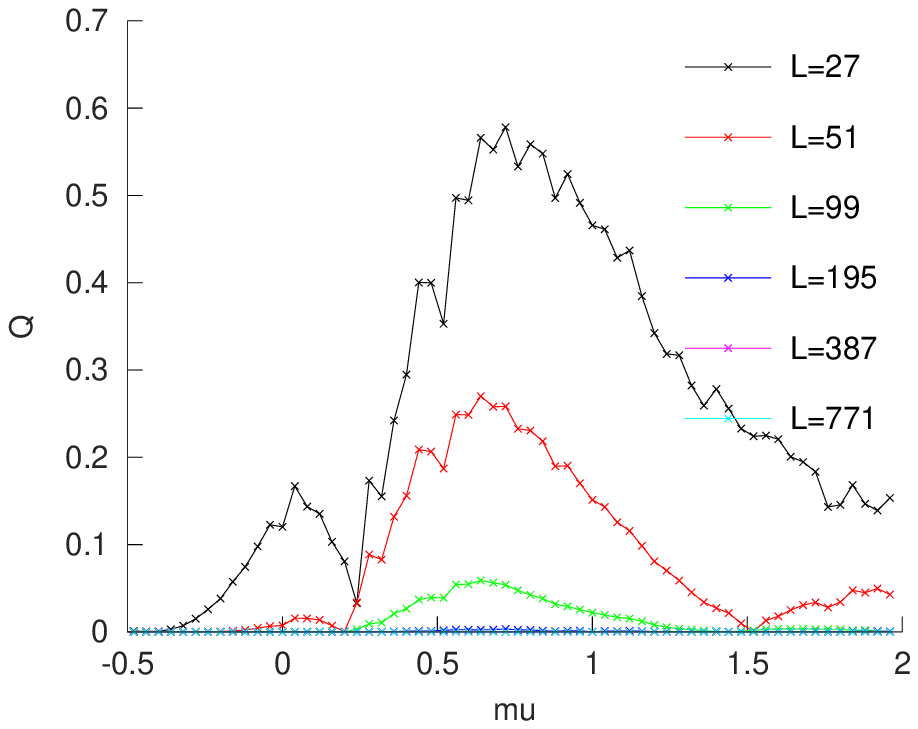}\quad
    \includegraphics[width=8cm]{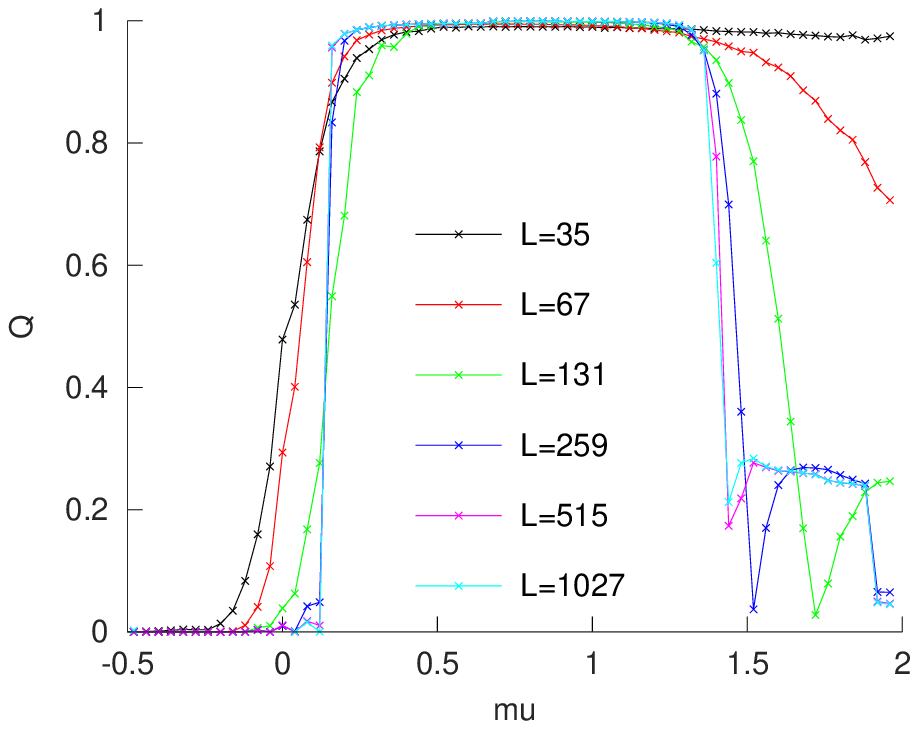}
    \caption{Order parameter of the nematic phase of the 6-mer model (top) and of
    the 8-mer model (bottom) versus the chemical potential $\mu$ per monomer. The data
    were computed using CTMRG with 648 states for the 6-mer model and 512 for the 8-mer model.
    The different curves correspond to different lattice sizes $L$ as indicated by the legend.
    }
  \label{fig7}
  \end{center}
\end{figure}

For comparison, the order parameter $Q$ of the 6-mer and 8-mer models are presented
on figure~\ref{fig7}. In the case of the 6-mer model, the order parameter vanishes as the
lattice size is increased indicating the absence of any intermediate nematic phase.
In contrast, in the case of the 8-mer model, the order parameter saturates over a wide range
of chemical potentials.

\subsection{Entropy of the central vertex}
\begin{figure}
  \begin{center}
    \psfrag{mu}[Bl][Bl][1][1]{$\mu$}
    \psfrag{S}[Bl][Bl][1][1]{$S$}
    \includegraphics[width=8cm]{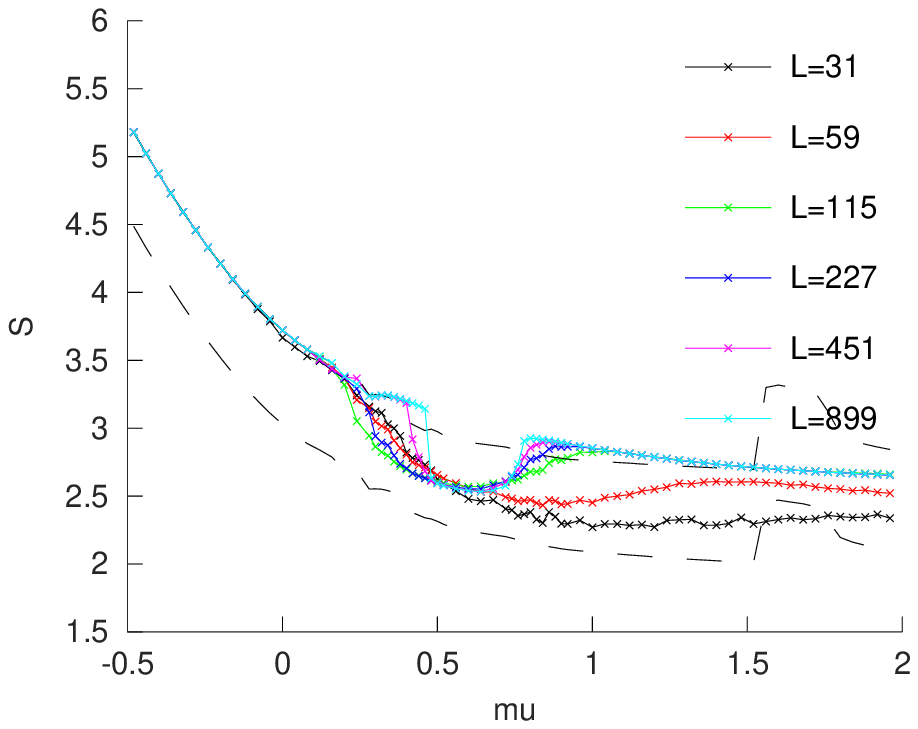}
    \llap{\raisebox{2.7cm}{\includegraphics[width=5.33cm]{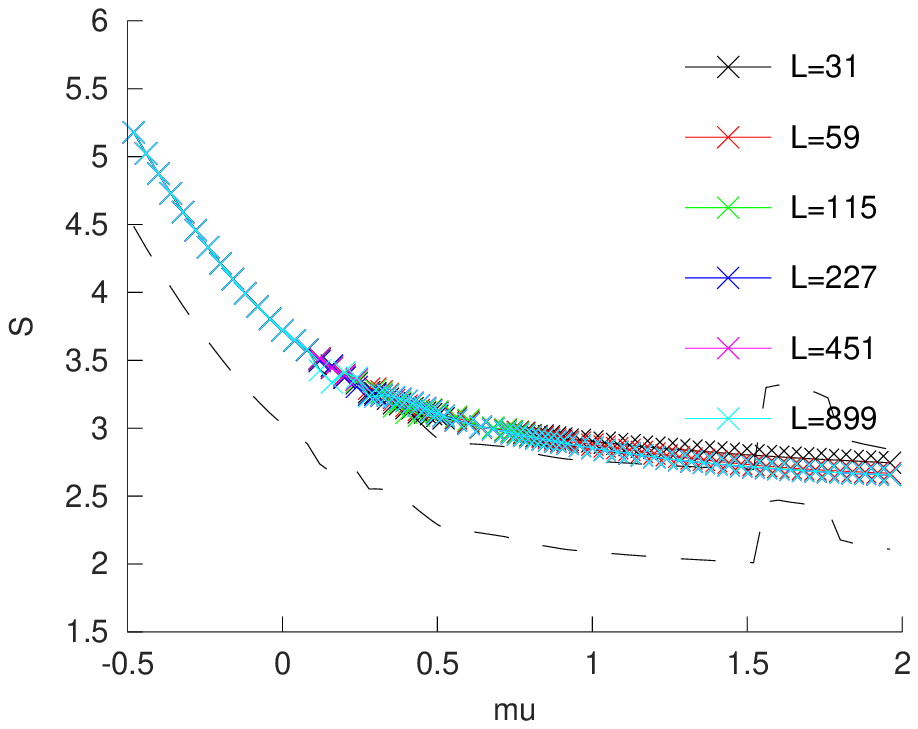}}}
    \quad
    \includegraphics[width=8cm]{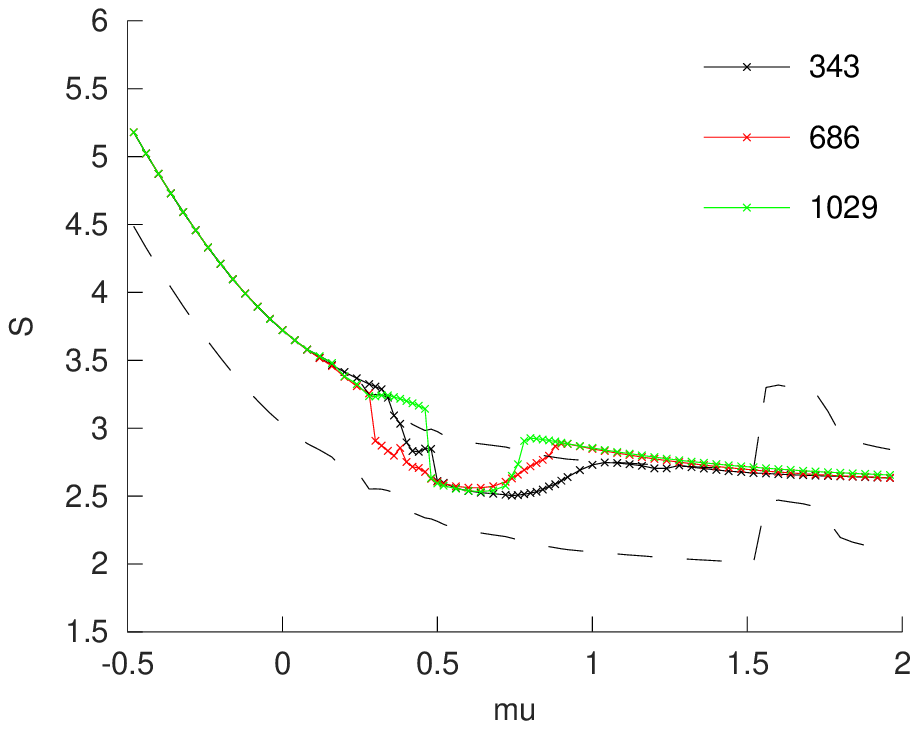}
    \llap{\raisebox{2.7cm}{\includegraphics[width=5.33cm]{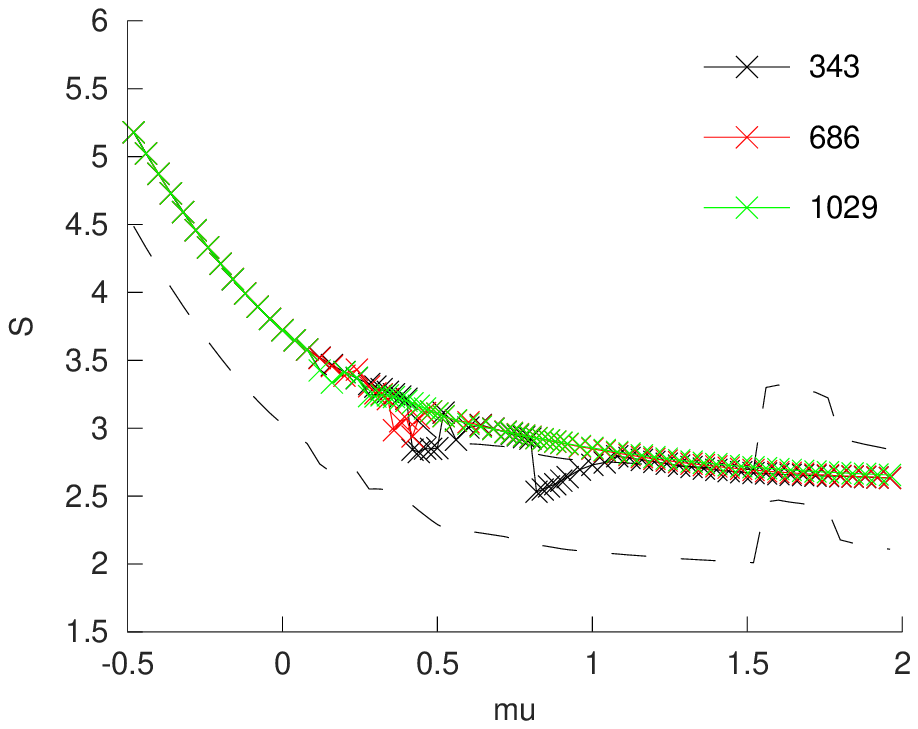}}}
    \caption{Entropy of the central vertex of the 7-mer model versus the chemical
    potential $\mu$ per monomer. On the top figure, the data were computed using CTMRG with 1029 states
    and the different curves correspond to different lattice sizes $L$ as indicated by the legend.
    On the bottom figure, the lattice size is fixed to $L=899$ but different numbers of states were
    kept in the CTMRG algorithm (343 in black, 686 in red and 1029 in green). In the inset, the
    same quantities are plotted but with OBCs instead of FBCs.
    The two ans\"atze (\ref{SOBC}) and (\ref{SFBC}) are plotted as dashed lines.}
  \label{fig8}
  \end{center}
\end{figure}

In the $k$-mer model, a single monomer lies on each site of the square lattice.
Therefore, the entropy of the central vertex is a local quantity, as the average density.
Considering the fact that a site may be empty with a probability $1-\langle n\rangle$ or
occupied by any of the $k$ possible monomers forming either a horizontal or a vertical $k$-mer
with a probability $\langle n\rangle/2k$, the entropy of the central vertex is
    \be S_{\rm OBC}=-(1-\langle n\rangle)\ln(1-\langle n\rangle)
    -2k\times {\langle n\rangle\over 2k}\ln {\langle n\rangle\over 2k}.
    \label{SOBC}\ee
In the nematic phase, the ${\mathbb Z}_2$ orientational symmetry between horizontal
and vertical $k$-mers is broken by FBCs and only one orientation of the $k$-mer is allowed.
The entropy of the central vertex is then expected to be
     \be S=-(1-\langle n\rangle)\ln(1-\langle n\rangle)
    -k\times {\langle n\rangle\over k}\ln {\langle n\rangle\over k}.\label{SFBC}\ee
The entropy per monomer $S/\langle n\rangle$ of the 7-mer model is plotted on
figure~\ref{fig8}. In the case of OBCs (presented in the inset), the curves are nicely
compatible with $(\ref{SOBC})$ for sufficiently large number of states. Finite-Size
corrections remain small.
With FBCs, a depletion appears in an intermediate range of chemical potentials. The entropy
of the central vertex is close to $(\ref{SFBC})$ in this depletion while it remains nicely
compatible with $(\ref{SOBC})$ outside. The existence of an intermediate nematic phase,
sensitive to the boundary conditions unlike the two disordered phases, explains the
numerical data. Like the order parameter, the depletion becomes thinner as the lattice
size or the number of states is increased. For 1029 states, the curves of the two largest
lattice sizes collapse between $\mu_1\simeq 0.47$ and $\mu_2\simeq 0.75$. The first
chemical potential is compatible with the location of the isotropic-nematic
transition as estimated above from the order parameter $Q$. The second is smaller
than the previous estimate but closer to the Monte Carlo estimate $\mu_2\simeq
0.795$~\cite{Kundu1}.

\begin{figure}
  \begin{center}
    \psfrag{mu}[Bl][Bl][1][1]{$\mu$}
    \psfrag{S}[Bl][Bl][1][1]{$S$}
    \includegraphics[width=8cm]{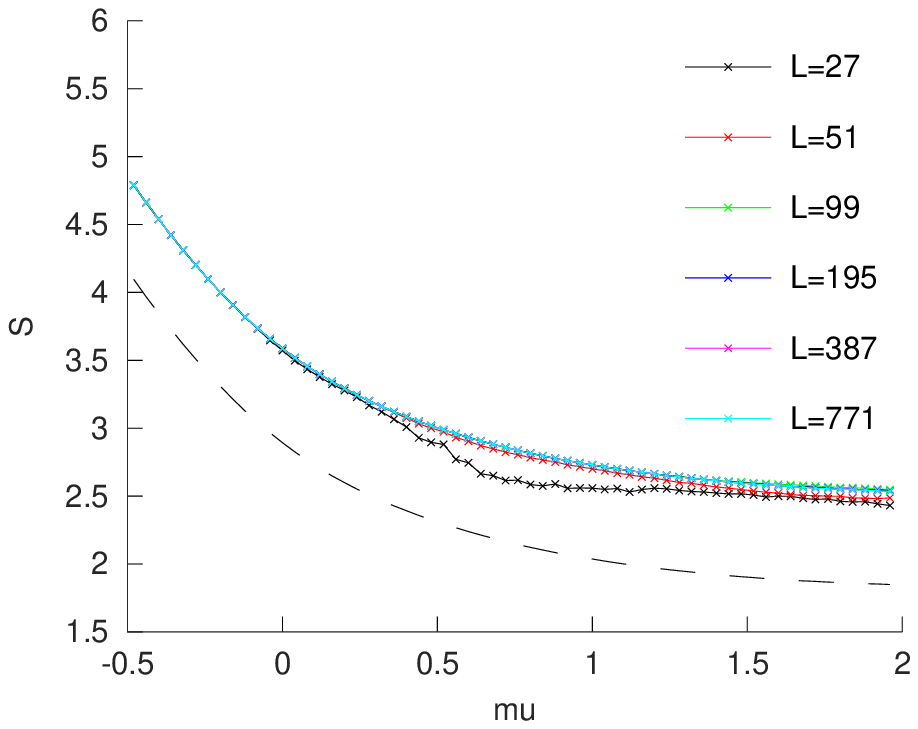}\quad
    \includegraphics[width=8cm]{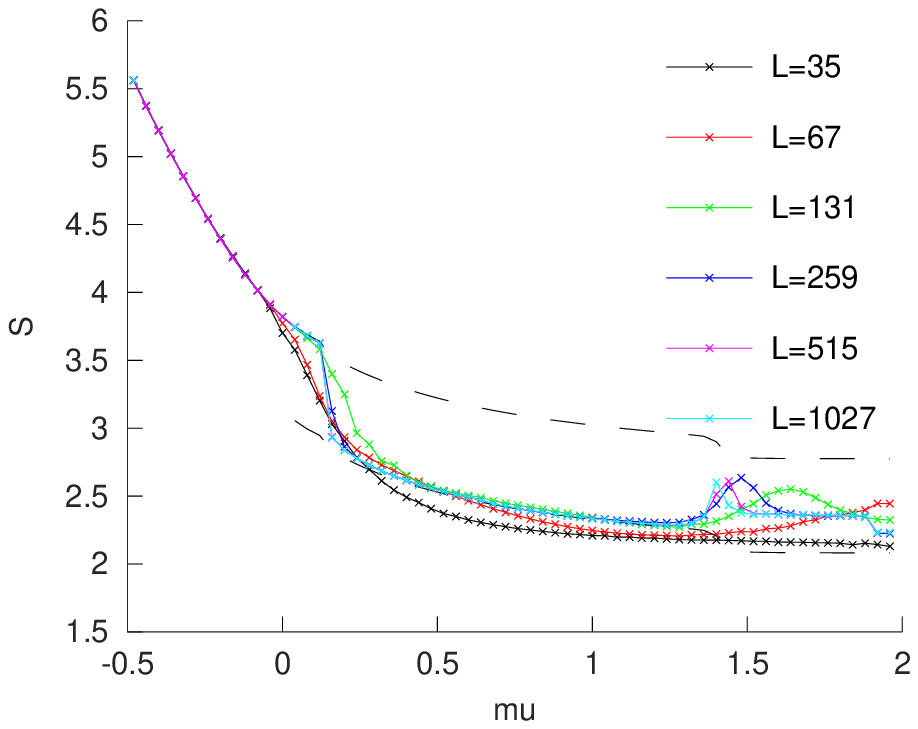}
    \caption{Entropy of the central vertex of the nematic phase of the 6-mer model (top)
    and of the 8-mer model (bottom) versus the chemical potential $\mu$ per monomer. The data
    were computed using CTMRG with 648 states for the 6-mer model and 512 for the 8-mer model.
    The different curves correspond to different lattice sizes $L$ as indicated by the legend.
    The two ans\"atze (\ref{SOBC}) and (\ref{SFBC}) are plotted as dashed lines using the numerical
    estimate of the density.}
  \label{fig9}
  \end{center}
\end{figure}

For comparison, the entropies of the central vertex for the 6-mer and 8-mer models are
presented on figures~\ref{fig9}. In the case of the 6-mer model, the entropy per
monomer $S/\langle n\rangle$ is compatible with $(\ref{SOBC})$ for all considered
chemical potentials. There is no signature of a nematic phase in this case. For the
8-mer model, the entropy $S$ displays a wide flat depletion where it is compatible with
$(\ref{SFBC})$. For large chemical potential, the entropy of the central vertex has not
returned to the value predicted by $(\ref{SOBC})$, as expected in the disordered phase.
It is not clear whether the transition nematic-isotropic is absent or the
number of states is still too small to achieve convergence.

\section{Entanglement entropies of the $k$-mer model}
\begin{figure}
  \begin{center}
    \psfrag{mu}[Bl][Bl][1][1]{$\mu$}
    \psfrag{Se}[Bl][Bl][1][1]{$S_A$}
    \includegraphics[width=8cm]{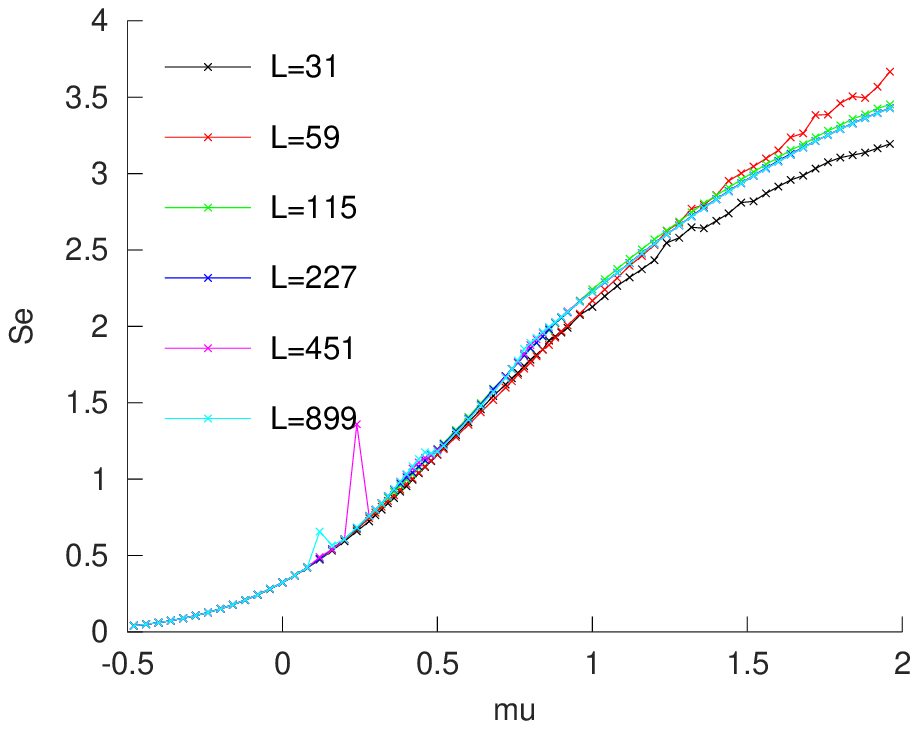}\quad
    \includegraphics[width=8cm]{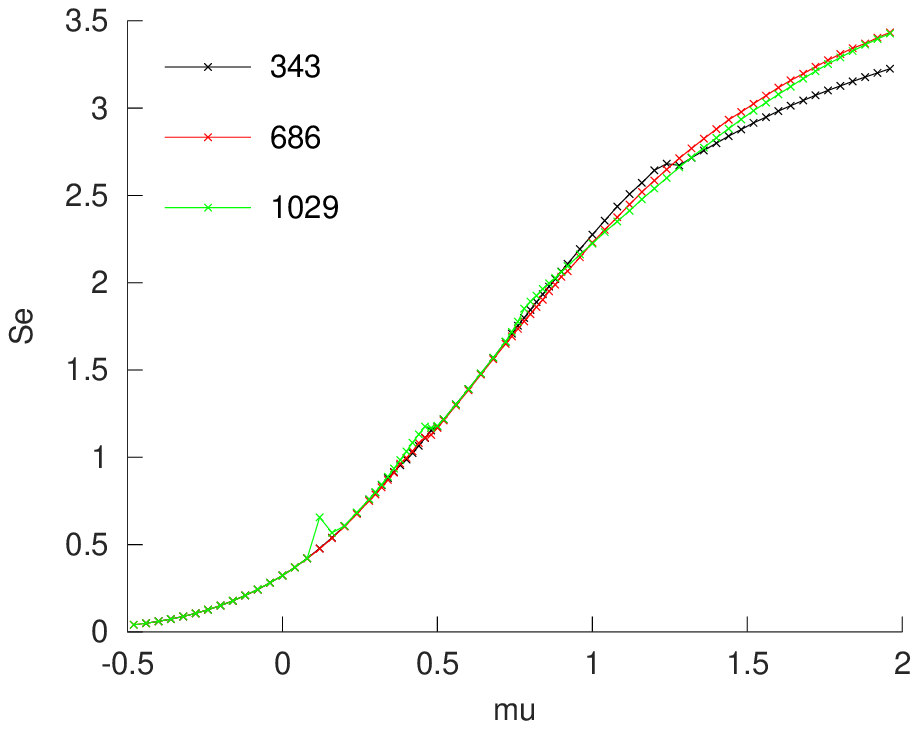}
    \caption{Von Neumann entanglement entropy of the 7-mer model versus the
    chemical potential $\mu$ per monomer. On the top figure, the data were computed
    using CTMRG with 1029 states and the different curves correspond to different
    lattice sizes $L$ as indicated by the legend. On the bottom figure, the lattice
    size is fixed to $L=899$ but different numbers of states were
    kept in the CTMRG algorithm (343 in black, 686 in red and 1029 in green).}
  \label{fig10}
  \end{center}
\end{figure}

\begin{figure}
  \begin{center}
    \psfrag{mu}[Bl][Bl][1][1]{$\mu$}
    \psfrag{Se}[Bl][Bl][1][1]{$S_A$}
    \includegraphics[width=8cm]{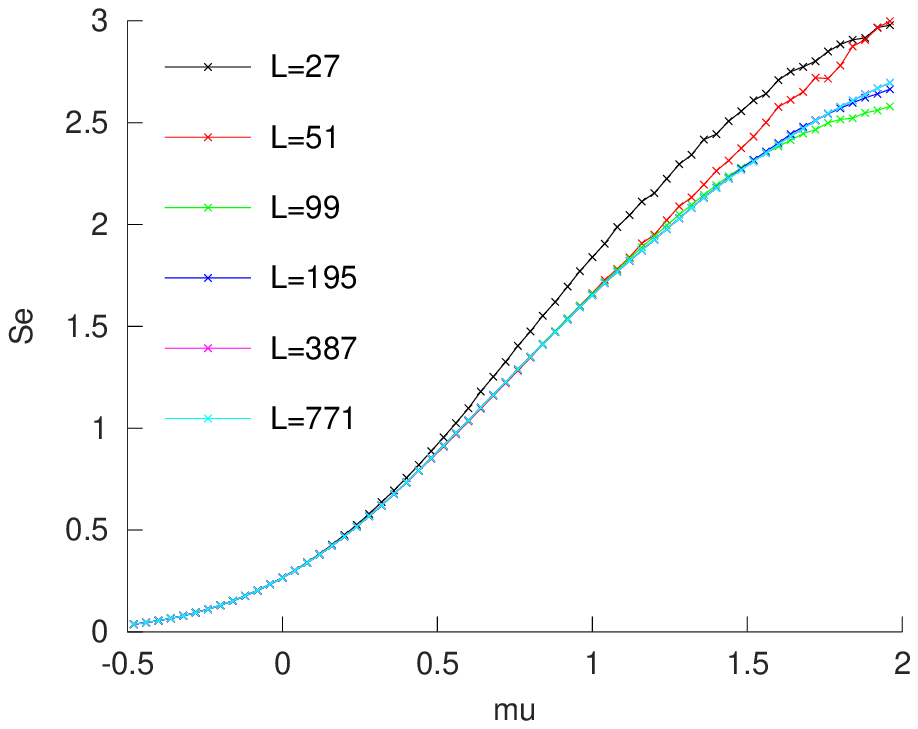}\quad
    \includegraphics[width=8cm]{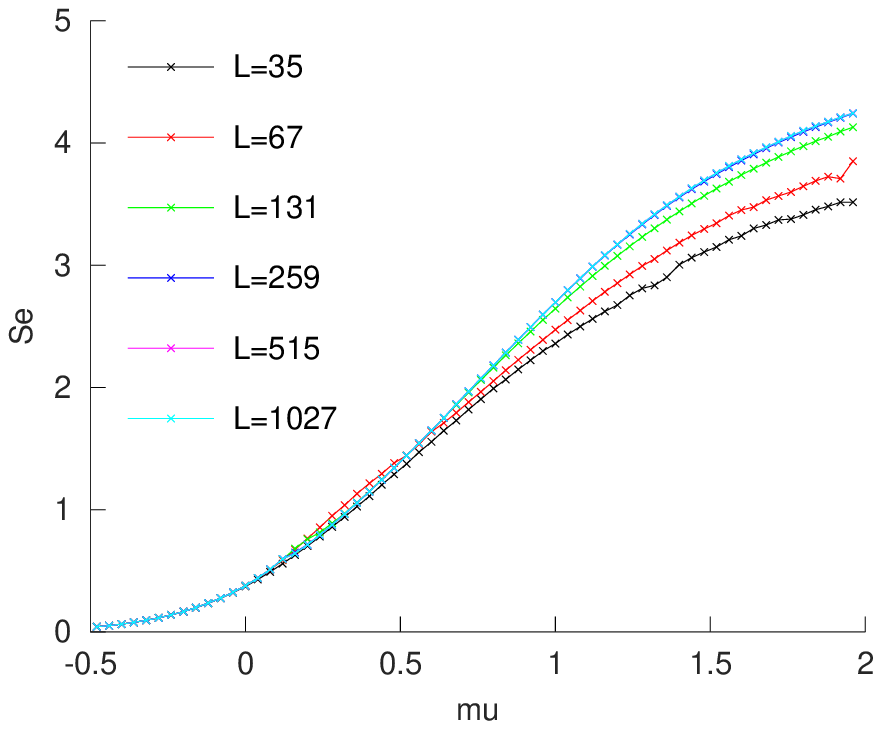}
    \caption{Entanglement entropy of the 6-mer model (top) and of
    the 8-mer model (bottom) versus the chemical potential $\mu$ per monomer. The data
    were computed using CTMRG with 648 states for the 6-mer model and 512 for the 8-mer model.
    The different curves correspond to different lattice sizes $L$ as indicated by the legend.
    }
  \label{fig11}
  \end{center}
\end{figure}

The von Neumann entanglement entropy is presented on figure~\ref{fig10} for the 7-mer model.
Apart from the two points out of the curve in the disordered phase, the entanglement entropy
grows monotonously with the chemical potential $\mu$. In particular, no peak is observed, even at large lattice
sizes, in apparent contradistinction with the assumption that the two transitions are continuous.
A small dependence with the lattice size is observed. However, this dependence is similar
to the one displayed by the average density (figure~\ref{fig5}) and differs from the
one of the 7-state clock model in its critical phase (figure~\ref{fig4}). Therefore, the usual
signature of a first, second order and even Berezinskii-Thouless transition is absent in
the 7-mer model. Increasing the number of states of the CTMRG algorithm does not change the
situation. In contrast, the entanglement entropy of a quantum state approximated by a MPS
is known to be bounded by a function of the logarithm of the dimension of the auxiliary
space~\cite{Verstraete,Schuch,Pollmann}. A relation similar to this one should hold for the truncated
corner transfer matrices. However, it can be observed on figure~\ref{fig10} that the
variation of the entanglement entropy is very small (and mostly negative!) when extending
the number of states from 686 to 1029. Therefore, the upper bound on the entanglement entropy
imposed by the number of states used in the calculations is not reached and it can be considered
that the estimated entanglement entropy has already reached its exact value with the considered
numbers of states. With OBC, the curve is very similar. Only finite-size corrections seems to
depend on boundary conditions. The curves are quite similar for the 6 and 8-mer models
(figure~\ref{fig11}). The entanglement entropy increases monotonously with the chemical potential.
Only the sign of the Finite-Size correction differs.

\begin{figure*}
  \begin{center}
    \psfrag{Lambda}[Bl][Bl][1][1]{$\rho_i$}
    \psfrag{n}[Bl][Bl][1][1]{$i$}
    \includegraphics[width=5cm]{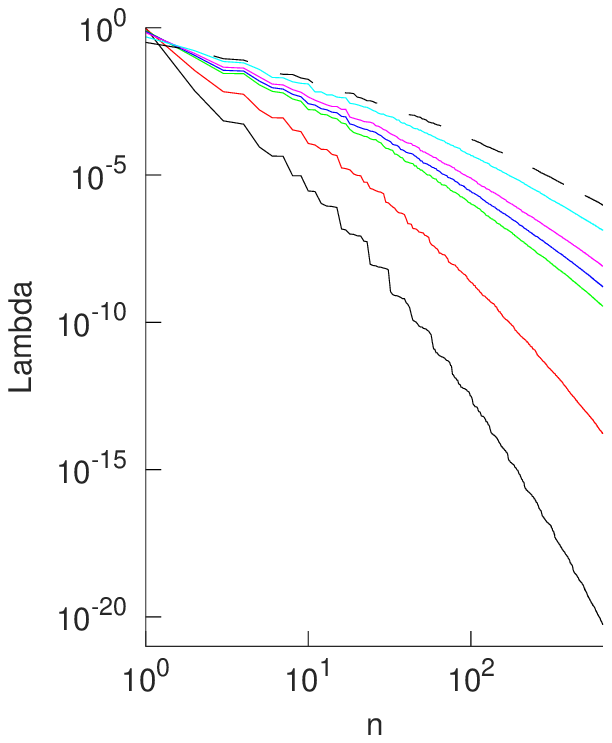}\quad
    \includegraphics[width=5cm]{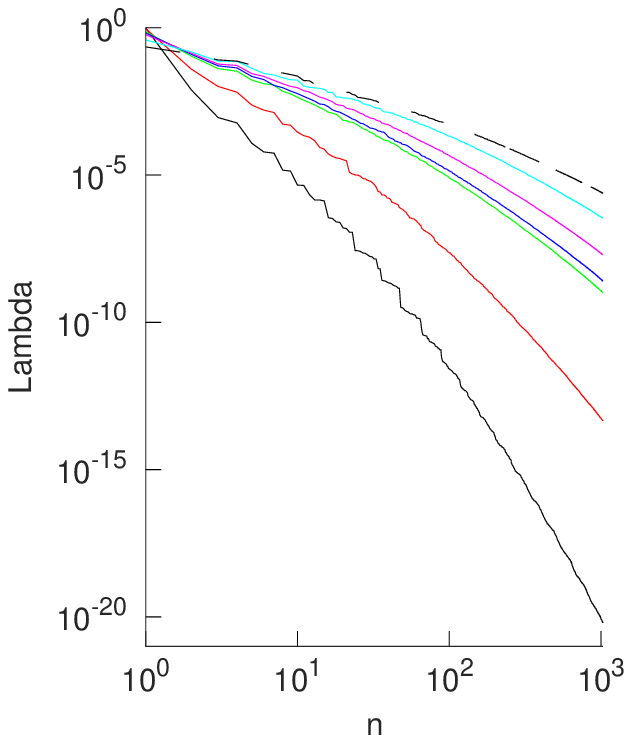}\quad
    \includegraphics[width=5cm]{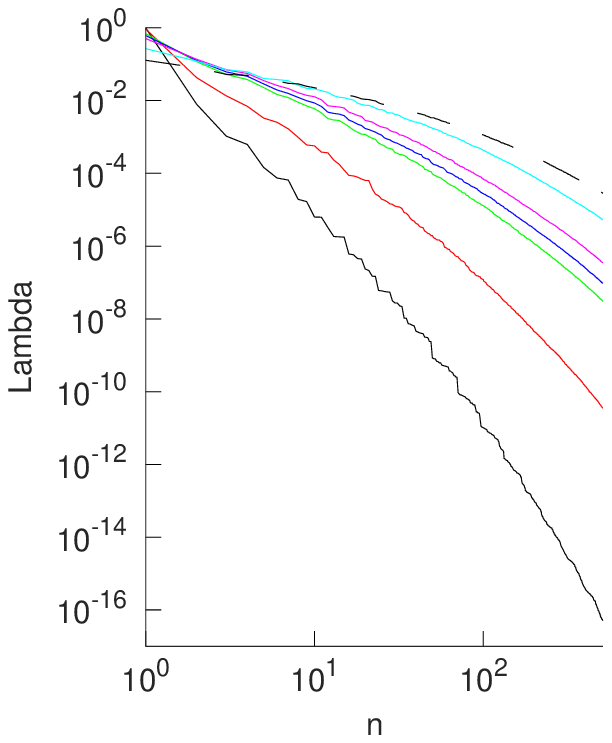}
    \caption{Entanglement spectrum of the 6 (left), 7 (center), and 8-mer (right) models
    at the largest lattice size and with the largest number of states considered. The different
    curves correspond to chemical potentials $-0.4$, $0$, $0.48$, $0.6$, $0.76$, $1.24$,
    $1.92$ (from bottom to top).}
   \label{fig12}
  \end{center}
\end{figure*}

The entanglement spectrum $\rho_i=\Lambda_i^4/\sum_i \Lambda_i^4$ is plotted on
figure~\ref{fig12} at different chemical potentials. No significant difference
between the 6, 7 and 8-mer model can be observed. The decay of $\rho_i$ with $i$
is slower than an exponential but faster than a simple power law. Moreover, the
decay becomes slower as the chemical potential is increased. In contrast, in lattice
spin models, the entanglement spectrum displays its slowest decay at the critical
point and the decay becomes faster as the temperature moves away from the
critical point.

\section{Conclusions}
The existence of two phase transitions for the $k$-mer model with $k\ge 7$, as previously
shown by means of Monte Carlo simulations, is confirmed by the study of the order parameter
and the entropy at the central vertex of a square lattice. {\CC However, the accuracy
reached by our CTMRG computations is not sufficient to determine the critical behavior
{\AG which is} associated to these transitions {\AG in order to} test the
{\AG conjectures} made from Monte Carlo simulations.
Nevertheless, in our CTMRG calculations the entanglement entropy increases
monotonously with the chemical potential. No peak is observed at the two transitions.}
In contrast, in the geometry considered in this work, the von Neumann and Renyi entanglement entropies of the $q$-state Potts model diverge as the logarithm of the lattice size
at the critical temperature when $q\le 4$ and is discontinuous
at the first-order transition when $q>4$.
{\CCb
We made CTMRG calculations of the Ising model with a number of states kept at each
truncation of the corner transfer matrix as small as 2 and of the 7-state Potts
model with only 7 states. The peak is shifted and rounded but is still present
and clearly visible. We infer that the absence of peak of the entanglement
entropy in the $k$-mer model cannot be explained by an insufficient number of
states in our CTMRG calculations.} Moreover, Conformal Field Theory predicts
that the entanglement entropy diverges in a universal manner with the lattice
size $L$, as ${c\over 6}\ln L$ for the von Neumann entanglement entropy and
${c\over 12}\left(1+{1\over n}\right)\ln L$ for the Renyi entropies (Fixed Boundary
Conditions). It is therefore inferred that the two transitions of the $k$-mer model
with $k\ge 7$ cannot be in the universality class of the Ising model, despite
the fact that a ${\mathbb{Z}}_2$ symmetry
is broken in the nematic phase. Moreover, if the transitions are continuous,
their long-distance behavior probably cannot be described by a Conformal
Field Theory.

\section*{Acknowledgements}
The numerical simulations were performed at the Explor meso-center
of the University of Lorraine. This research was partially supported by
APVV-16-0186 (EXSES).
\\

The two authors made an equal contribution to the paper.

\end{document}